# Beyond Simple Structure-Function Relationships: The Interplay of Geometry, Electronic Structure and Molecule/Electrode Coupling in Single Molecule Junctions

Nathan D. Bamberger[1], Dylan Dyer[1], Keshaba N. Parida[1], Tarek H. El Assaad,[1] Dawson Pursell[1], Dominic V. McGrath[1], Manuel Smeu[2] and Oliver L.A. Monti[1,3,*]

[1]Department of Chemistry and Biochemistry, University of Arizona, 1306 E. University Blvd., Tucson, Arizona 85721, USA

[2]Department of Physics, Binghamton University – SUNY, 4400 Vestal Parkway East, Binghamton, New York, 13902, USA

[3]Department of Physics, University of Arizona, 1118 E. Fourth Street, Tucson, Arizona 85721, USA

ABSTRACT:

Structure-function relationships constitute an important tool to investigate the fundamental principles of molecular electronics. Most commonly, this involves identifying a potentially important molecular structural element, followed by designing and synthesizing a set of related organic molecules, and finally interpretation of their experimental and/or computational quantum transport properties in the light of this structural element. Though this has been extremely powerful in many instances, we demonstrate here the common need for more nuanced relationships even for relatively simple structures, using both experimental and computational results for a series of stilbene derivatives as a case study. In particular, we show that the presence of multiple competing and subtle structural factors can combine in unexpected ways to control quantum transport in these molecules. Our results clarify the reasons for previous widely varying and often contradictory reports on charge-transport in stilbene derivatives, and highlight the need for refined multidimensional structure-property relationships in single molecule electronics.

# 1. Introduction

The field of molecular electronics—which involves building circuits incorporating individual small organic molecules as components[1]—has the potential to enable smaller, more efficient, and more flexible electronic devices in the future,[2] and also serves as a useful test bed for fundamental investigations of quantum transport and molecular physical chemistry.[3] In order to advance both of these goals, predictive structure-property relationships are needed that allow transport properties to be intentionally controlled via molecular design. This widely embraced goal in the molecular electronics research community usually proceeds by identifying a potentially important design element in a molecular structure, followed by testing this prediction with a series of structurally related molecules that seemingly isolate the importance of this particular element.[4–8] While this often constitutes a powerful approach, it ignores the potential for interaction – deliberate or undesired – between multiple coupled degrees of freedom for molecules in a junction. As a consequence, seemingly minor modifications to the molecular structure may result in widely different transport behavior, confounding the interpretation in terms of underlying principles.

One such case where this is quite evident are cis *vs*. trans isomers of stilbene-like moieties,[9–13] a structural element that commonly appears in molecular scaffolds considered for electronics applications, especially in the context of photo-activated switches.[14,15] Single molecule conductance—the most fundamental transport property—has been investigated both theoretically and experimentally for a stilbene-like molecule,[16] as well as for several derivatives of the structurally similar azobenzene moiety.[17–22] However, such studies have produced results varying from an ~2× higher conductance for cis over trans,[16] to the exact opposite,[21] and even all the way up to ~100× higher conductance for trans[18] or ~30× higher conductance for cis.[22] Part of the reason for this inconsistency is likely that changing from trans to cis unavoidably causes multiple structural changes at once, each of which can affect conductance (Figure 1): The length of the molecule decreases, which is expected to increase conductance by reducing the tunneling barrier width;[23] steric hindrance causes each ring in the cis molecule to twist out of plane by ~35º, which is expected to decrease conductance by partially breaking the conjugated π-system;[24–26] the electronic structure of cis and trans isomers and their respective energy level alignments with the electrodes may differ; and the modified geometry of the molecule, especially the angles at which linkers attached in the standard para positions extend out from the backbone, may impact the orbital alignment between metal and molecule and hence the efficiency of electronic coupling between the two. Attaching linkers in positions other than para often produces destructive quantum interference,[27] which would add yet another confounding factor influencing conductance. Therefore, while most previously developed structure-property relationships focus on one design property at a time,[6,24,28–30] the effects of cis/trans isomerization on single-molecule conductance cannot be expected to be as simple. Instead, a more nuanced structure-property relationship is required that considers the roles of multiple effects simultaneously, especially interactions *between* them.

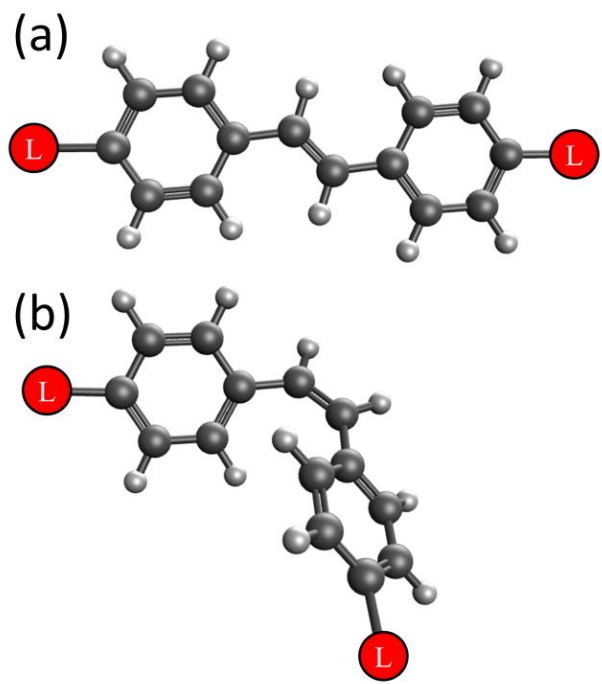

Figure 1. Generic structures of trans (a) and cis (b) stilbene derivatives with para-connected linker groups (represented by red circles), illustrating three key differences that arise upon isomerization: 1) The linker-to-linker length decreases by ~25%; 2) the second ring twists out of plane with respect to the first; and 3) the two linkers are no longer pointed 180° opposite each other.

In particular, previous results have suggested that differences in how cis and trans geometries influence molecule/metal coupling may play an important role in determining single-molecule conductance. For example, calculations by Osella et al.[19] suggested that chemisorption vs. physisorption produce opposite cis/trans conductance orderings, and Martin et al.[16] rationalized their experimental results by positing that the cis geometry leads to stronger coupling by allowing not just the linker groups but also one phenyl ring to come into close contact with the metal electrode. Nevertheless, these effects remain poorly understood. One especially important case that has not yet been explored is how the coupling of conformationally flexible linker groups to the metal electrodes might be differentially affected by cis vs. trans backbone geometries. This possibility is motivated by density functional theory (DFT) calculations for the common -SMe (methylsulfide) linker group, which have shown that the barrier to rotation of this linker is quite small,[31,32] while at the same time molecule/metal electronic coupling, and hence conductance, is strongly controlled by the angle between the S-Au bond and the π-system, with maximal coupling occurring for a 90° angle.[31,33] Any differences in the geometric constraints that cis and trans geometries place on -SMe linker orientation therefore have the potential to be a major factor by which cis/trans isomerization impacts single-molecule conductance, and such interactions likely apply more generally to other flexible linkers as well.

In order to investigate this possibility and explore the interactions between cis/trans isomerization and molecular conductance more generally, in this work we use a combination of experiment and theory to investigate the single-molecule conductance of a series of custom-designed stilbene derivatives functionalized with -SMe linkers (Scheme 1). This series includes the cis and trans versions of this structure (trans-OPV2-2SMe and cis-OPV2-2SMe), as well as single-linker versions (trans-OPV2-1SMe and cis-OPV2-1SMe) to help understand the role of -SMe/metal orbital coupling. In order to help elucidate and control for potential confounding effects from molecular twisting that both breaks the π-system conjugation and potentially forces new linker/electrode geometries, we also include a version of the cis molecule with the rings locked into a fully planar geometry (L-cis-OPV2-2SMe) and a version of the molecule with the π-conjugation fully broken by saturating the double-bond linkage (sat-OPV2-2SMe). Together, this series thus explores and helps disentangle multiple key design parameters used throughout the single molecule literature, and sheds light on interactions between them that determine molecular conductance. Our study therefore helps clarify the reasons for the widely differing reported conductance values for seemingly similar molecular structures with cis vs. trans double bonds.

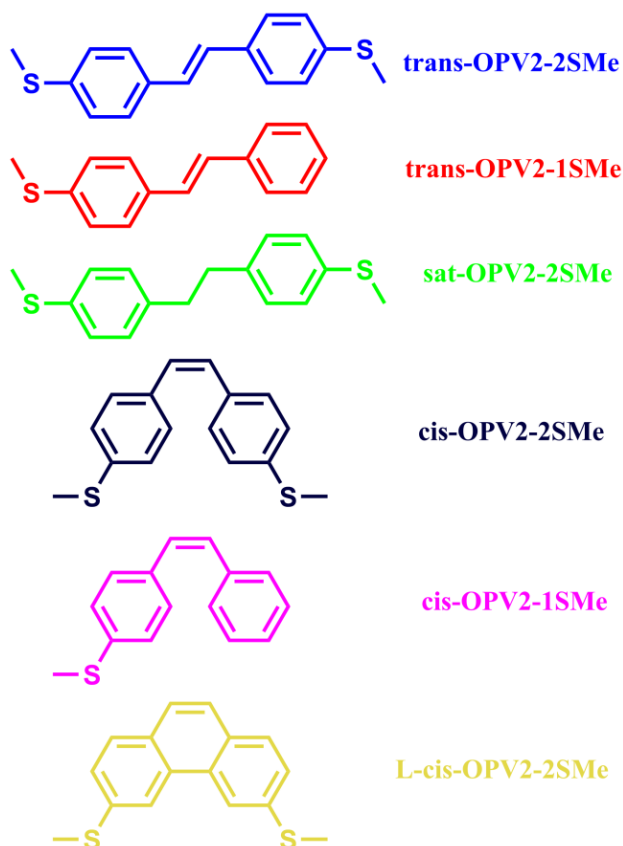

Scheme 1. Structures of the six OPV2 molecules considered in this work, as well as the naming and color conventions used throughout.

In the remainder of this paper, we present experimental evidence that the geometric constraints of the cis geometry result in less efficient metal/molecule electronic coupling, and hence lower conductance, compared to the trans geometry. We also show how this interaction between backbone conformation and linker orientation can explain the striking observation that cis-OPV2-2SMe, cis-OPV2-1SMe, and L-cis-OPV2-2SMe all display essentially the same peak conductance value. We then use computational results to support these interpretations and offer additional insight into how -SMe linker orientation affects single-molecule conductance. These findings illustrate the limitations of "divide and conquer" structure-function relationships in systems where multiple effects interact, and begin to develop more nuanced relationships that will help the readily accessible cis/trans structural motif to be predictively used to control conductance. In addition, these results highlight the fact that the potential complexity introduced by -SMe linkers—and likely other common linkers with conformational flexibility as well—must be taken into account when designing molecules for transport experiments.

## 2. Methods

### 2.1 Experimental Methods

The experimental conductance of each OPV2 molecule from Scheme 1 was measured using a custom-built mechanically controlled break junction (MCBJ) set-up described previously.[4,34,35] Briefly, MCBJ samples were fabricated on a phosphor bronze substrate coated with an insulating layer of polyimide. A pattern containing a thin (~100 nm) gold constriction was defined with electron beam lithography and then coated with 4 nm of titanium and 80 nm of gold. Reactive ion etching with an $O_2$/$CHF_3$ plasma was used to under-etch the polyimide and create an ~1 µm long free-standing gold bridge.

To collect breaking traces, each sample was clamped into a custom three-point bending apparatus with a push rod controlled by both a stepper motor (for coarse movements) and a 40 µm piezo actuator (for movement during trace collection). Conductance through the gold bridge or nanogap was measured at 20 kHz while applying an 100 mV bias, using a custom high-bandwidth Wheatstone bridge amplifier.[36] For each trace, coarse movements were used to achieve a junction conductance between 5 and 7 $G_0$ (1 $G_0$ = 77.48 µS,[37] the quantum of conductance), then the piezo was extended at 60 µm/s to break the bridge while recording trace data. Thousands of traces were collected for each sample using a custom LabVIEW program.

All molecules were synthesized on site and characterized by NMR and mass spectrometry (SI section S.1). Molecular solutions of 1 and/or 10 µM concentration in HPLC grade (>99.7%, Alfa Aesar) dichloromethane (DCM) were created for deposition on MCBJ samples. For each sample, 2000+ consecutive breaking traces were collected after the deposition of pure DCM as a negative control and to determine a junction attenuation ratio.[34] Next, the molecular solution was deposited inside a Kalrez gasket placed on the center of the sample using a clean glass syringe. Following the approach of Bamberger et al.,[34] multiple datasets were collected using each sample, and datasets from at least two distinct samples were collected for each molecule (see SI section S.2 for details on all datasets used in this work).

The molecular features observed for the OPV2 series often have complex shapes due to convolution with the tunneling background signal, which additionally is known to vary between datasets.[34,38] A segment clustering tool described previously[34] was thus used to unambiguously identify a "main plateau cluster" from each analyzed dataset. The conductance distribution of trace segments assigned to each plateau cluster was then fit with a single unrestricted Gaussian to determine a peak conductance value. Following the method of Bamberger et al.,[34] each dataset was clustered 12 different times to account for uncertainty in the optimal value of the *minPts* clustering parameter (see SI section S.3 for details).

### 2.2 Computational Methods

The molecules of interest were first relaxed as isolated species with the Orca code[39] using the B3LYP[40,41] functional with the 6-311++G(d,p) basis set. The relaxed molecules were then placed between Au electrodes that taper to a single Au atom which interacts directly with the lone pair of the -SMe group (see Figure S16a). For the single-linker versions, the Au atom was arranged to interact directly with the π-system of the benzene ring without the -SMe group (see Figure S16b).

A series of structure relaxations were carried out in the Vienna Ab inito Simulation Package (VASP)[42,43] for each molecule to determine the lowest-energy electrode-electrode separation. We employed the Perdew-Burke-Ernzerhof (PBE)[44] functional with the DFT-D3 method of Grimme[45] to account for van der Waals interactions. The plane wave cutoff was set to 400 eV and relaxations continued until residual forces on atoms allowed to relax were lower than 0.02 eV/Å. The optimized molecular junction was then built into a two-probe geometry by extending the Au electrodes (see Figure S16c).

Electron transport was calculated through the optimized two-probe geometries using the non-equilibrium Green's function technique combined with density functional theory (NEGF-DFT) as implemented in the Nanodcal code.[46,47] The NEGF-DFT approach uses the retarded Green's function to obtain the transmission function, $T(E)$, which represents the probability that an electron with a given energy $E$ is transmitted from the left electrode through the molecule into the right electrode. Integrating the transmission function over a specific energy range gives the current for the corresponding bias window. In the limit of the bias voltage approaching zero, the transmission at the Fermi energy gives the low-bias conductance in units of $G_0$. Details of this approach have been provided in our previous publications.[4,48]

## 3. Results and Discussion

### 3.1 Experimental Results

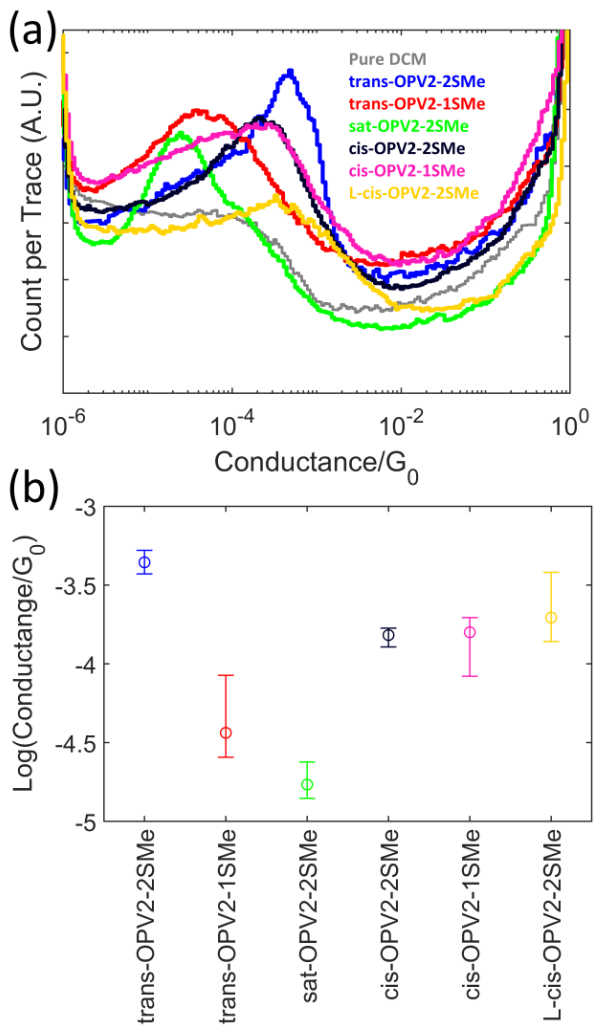

Figure 2. (a) Overlaid example 1D conductance histograms for each OPV2 molecule and a negative control with only DCM. (b) Summary of the peak conductance values determined for each molecule, across multiple datasets, using segment clustering.

Representative one-dimensional (1D) conductance histograms for each OPV2 molecule, as well as for pure DCM, are overlaid in Figure 2a. Each of these histograms shows a broad peak whose location is largely reproducible across the other datasets collected with the same molecule (see SI section S.3 for further details). Moreover, the main plateau features identified in each dataset by segment clustering tend to agree well with these peaks, indicating that segment clustering is reliably extracting molecular signatures (SI section S.4). The segment clustering results are summarized in Figure 2b: Each point represents the median from among the set of all 12 peak conductance values produced by segment clustering from all datasets collected with each molecule, and the error bars represent the range of the middle 67% of those values.[4] These error bars are thus a measure of the uncertainty due to both dataset-to-dataset variation and ambiguity about the exact cluster bounds, but should not be interpreted as a single universal measure of experimental error.

Many of the relative conductance relationships in Figure 2b are in agreement with the predictions of "standard" single-dimensional structure-property rules. For example, going from trans-OPV2-2SMe to sat-OPV2-2SMe, which fully breaks the conjugated π-system while leaving the molecular length largely unchanged, causes the conductance to drop by over an order of magnitude, which is consistent with both previous measurements[10] and the widely accepted principle that conjugated π-systems extending from linker to linker are needed to produce highly-conducting molecules.[5,49] Similarly, multiple studies[13,50–52] have found that single-linker molecules (including trans-OPV2 with single -SMe linkers)[13] conduct significantly less than double-linker molecules, and this agrees with our results for trans-OPV2-1SMe vs. trans-OPV2-2SMe. Either direct gold-π coupling and/or two-molecule junctions involving π-π stacking have been proposed to explain the lower conductance of single-linker molecules,[13,50,52] and both structures could arise in our experiments. The apparent lengths of our molecular features suggest, however, that direct gold-π coupling is the more likely explanation (SI section S.5).

Strikingly, Figure 2b also reveals the surprising result that approximately the same conductance was measured for all three cis molecules, with all three *less* conductive than trans-OPV2-2SMe. This finding does not conform to "standard" structure-property rules: these might expect the shorter cis molecules to conduct *more* than the trans; they would further predict cis-OPV2-1SMe to have significantly lower conductance than cis-OPV2-2SMe, in analogy to the trans structures; and they would predict L-cis-OPV2-2SMe to have a conductance perhaps as high or higher than trans-OPV2-2SMe due to full conjugation combined with a shorter length. A new, more nuanced structure-property relationship is thus required to explain these results. In particular, the fact that the -2SMe and -1SMe molecules have very different conductances in a trans geometry but similar conductances in a cis geometry suggests that the role of the identical -SMe linker groups must depend on the backbone geometry in some way. We thus next examine the role of those linkers in detail, and uncover how they can explain the surprising conductances of each of the three cis molecules.

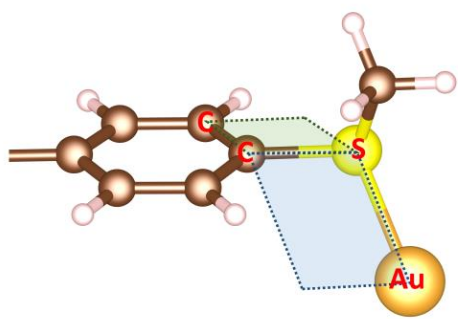

Figure 3. Definition of the Au-S-C-C dihedral angle relevant for molecule-electrode coupling.

It has been previously shown by DFT that, for conjugated molecules with -SMe linkers, the orientation of the Au-S bond relative to the π-system—quantified by the Au-S-C-C dihedral angle shown in Figure 3—strongly controls the strength of metal/molecule electronic coupling, and hence conductance. Coupling is maximized at an angle of 90º, which allows the sulfur lone pair involved in the Au-S bond to align with the extended π-system of the molecule, and is minimized at 0º.[31,33] In light of this, we hypothesize that the geometric constraints imposed by the cis-OPV2 backbone force the Au-S-C-C angles to spend more time farther from 90º, while in the less-constrained trans-OPV2 geometry these angles are free to spend more time closer to the optimal 90º. As a result, the average metal/molecule electronic coupling in cis-OPV2-2SMe and L-cis-OPV2-2SMe would be weaker than in trans-OPV2-2SMe, explaining the lower conductances we measured for these cis molecules despite their shorter length.

The fact that, relative to trans-OPV2-2SMe, conductance is lowered to a similar extent in both cis-OPV2-2SMe and L-cis-OPV2-2SMe, despite the absence of inter-ring twisting in the locked version, suggests that the weaker metal/molecule electronic coupling caused by the cis geometry is the more important effect in this series of structures. That is not to say that inter-ring twisting has no effect on conductance: the central estimate of conductance for L-cis-OPV2-2SMe in Figure 2b is ~1.3× higher than for cis-OPV2-2SMe, on the same order as the factor of ~2 predicted by the well-known "cos$^2$ rule" for twisting in conjugated systems[24–26] (though the difference in Figure 2b is small compared to the uncertainties in each measurement). However, even with this modest increase in conductance for the planar locked molecule, L-cis-OPV2-2SMe remains about two times *less* conductive than the *longer* trans-OPV2-2SMe. We thus conclude that weaker metal/linker electronic coupling is the dominant way in which the cis geometry influences conductance in these structures.

Finally, to explain the near-identical conductances of cis-OPV2-2SMe and cis-OPV2-1SMe, we propose that the *overall* metal/molecule coupling is quite similar between the two molecules. In particular, we propose that in cis-OPV2-1SMe, the lack of a second point for covalent, mechanical coupling between molecule and electrodes likely results in less geometric constraint on the remaining Au-S-C-C angle. Therefore, we suggest that cis-OPV2-2SMe has medium-to-weak, constrained -SMe/Au coupling on both sides, whereas cis-OPV2-1SMe has strong, unconstrained -SMe/Au coupling on one side but weak π/Au coupling on the other, averaging out to produce similar conductance values.

Previous studies considering the role of -SMe linker orientation on single-molecule conductance have focused on how the range of orientations that might be probed by experiment can help explain the breadth of measured conductance distributions or the overall conductance relative to other linker types.[31–33,53,54] The results in this work thus constitute the first experimental evidence that -SMe linker orientation interacts with other structural attributes such as cis vs. trans backbone geometry. This type of interaction is an important consideration when using -SMe or other linkers with multiple conformational degrees of freedom, and demonstrates that more nuanced structure-property relationships are needed in such cases where multiple structural effects are in play at the same time.

## 3.2 Computational Results

To further explore the interaction between cis/trans geometry and -SMe/Au coupling efficiency, we carried out NEGF-DFT calculations for the four two-linker OPV2 molecules from Scheme 1. In the case of the two single-linker molecules, calculations may not be able to offer the same level of reliable insights into transport through these structures because of the much larger conformational space available to single-linker molecules, combined with significant ambiguity about how such molecules bind in experimental junctions (see SI section S.7 for further details on these sources of ambiguity). In the following discussion, we thus focus exclusively on the computational results for the two-linker molecular structures.

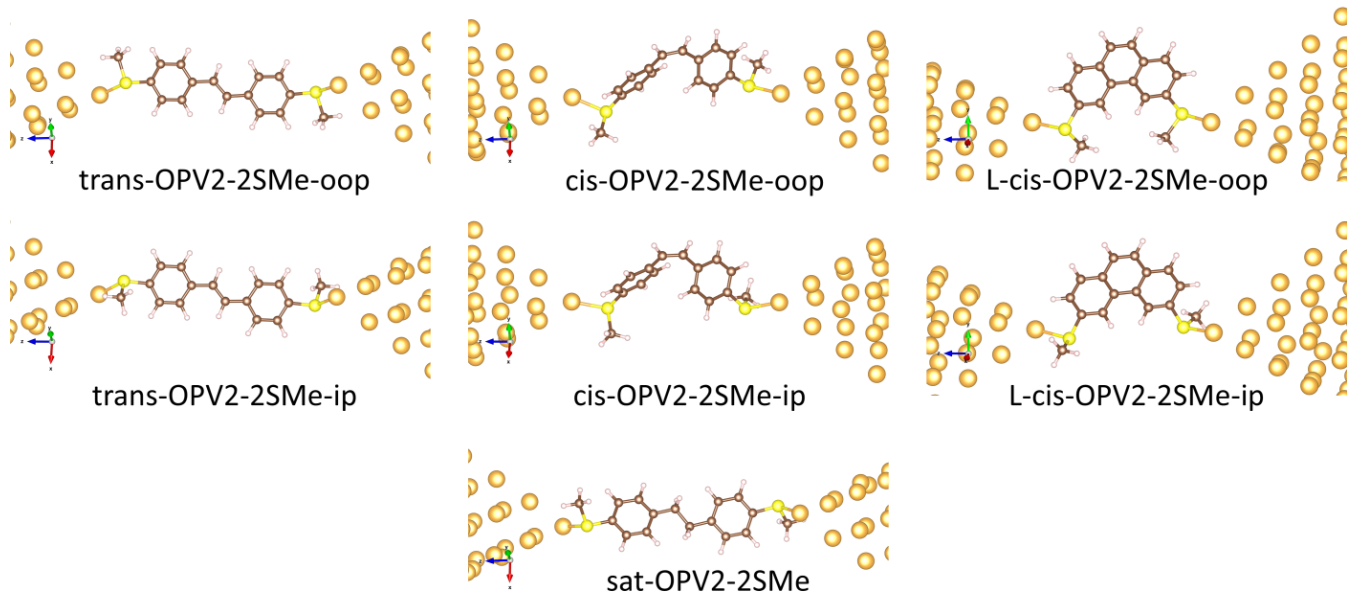

Figure 4. The relaxed structures of each two-linker OPV2 molecule used for NEGF-DFT transport calculations. For trans-OPV2-2SMe, cis-OPV2-2SMe, and L-cis-OPV2-2SMe, two relaxed conformations were generated by starting both S-Au bonds either in-plane (ip) or out-of-plane (oop) with respect to their attached rings, and then minimizing the energy. For sat-OPV2-2SMe, only a single conformation was considered because saturating the double bond has a much larger effect on conductance than linker orientation.

In order to approximately span the conformational space available to their -SMe linkers, for each of trans-OPV2-2SMe, cis-OPV2-2SMe, and L-cis-OPV2-2SMe we rotated the methyl groups to create two different junction conformations with both gold-sulfur bonds either primarily in-plane (ip) or out-of-plane (oop) with respect to the phenyl rings, and then allowing the geometry to fully relax from each of those starting points. These two relaxed conformations, which we label ip and oop, thus help us consider the two extremes of high- or low-coupling efficiency for each molecule in physically reasonable geometries (Figure 4). This is necessarily an incomplete description of the experimentally probed junctions, which likely sample a large number of conformations (see, e.g., SI section S.8), and in which variations in the atomic structure of the electrodes may also affect the -SMe orientations. Nevertheless, these conformations roughly span the available space and aid in interpreting our experimental results. Note that quantitative agreement between NEGF-DFT and experimental results is not expected in any case due to, among other things, the well-known tendency of DFT to underestimate band-gaps.[55,56] For sat-OPV2-2SMe, only a single conformation is considered since the saturated backbone, rather than linker orientation, dominates molecular conductance.

As shown in Table 1, the Au-S-S-C dihedral angles in both cis molecules are significantly farther from the optimal coupling angle of 90º compared to trans-OPV2-2SMe, in either the ip or oop conformations. This is consistent with our hypothesis, based on the experimental results, that a cis backbone imposes geometric constraints that force the -SMe linkers to spend more time in positions that experience weaker metal/molecule electronic coupling. The effect of these deviations on molecular conductance can be approximated by multiplying the square of the sine of each dihedral angle (maximal for $D1 = D2 = 90°$), which shows that the differences observed in Figure 4 have a significant impact on conductance (Table 1). Comparison with experimental results shows that this approximation captures the main qualitative differences in conductances between trans-, cis- and L-cis-OPV2-2SMe quite well.

Table 1. The two Au-S-C-C dihedral angles (D1 and D2) for the relaxed structures in Figure 4, shaded according to whether the S-Au bonds started from in-plane (ip) or out-of-plane (oop) positions prior to relaxation. Also included is the product of the sine squared of both angles as an approximation of the expected drop in conductance due to deviation from the ideal 90º.

| Molecule | D1 | D2 | $\sin^2(D1)\sin^2(D2)$ |
|---|---|---|---|
| trans-OPV2-2SMe-oop | 97.1º | 102.3º | 0.94 |
| cis-OPV2-2SMe-oop | 72.5º | 63.9º | 0.73 |
| L-cis-OPV2-2SMe-oop | 60.1º | 51.9º | 0.47 |
| trans-OPV2-2SMe-ip | 65.0º | 62.1º | 0.64 |
| cis-OPV2-2SMe-ip | 45.1º | 38.1º | 0.19 |
| L-cis-OPV2-2SMe-ip | 47.1º | 38.5º | 0.21 |

The transmission functions for each of the conformations in Figure 4 are shown in Figure 5. The Fermi energy ($E_F$) in these calculations falls very close to the LUMO-like transmission peak, but NEGF-DFT often misplaces $E_F$ within the HOMO-LUMO gap,[30,57] and the orbital composition of the -SMe linker is expected to induce HOMO-dominated transport.[5] We therefore focus on the region containing the transmission function minima and the trailing edge of the HOMO peaks (-0.75 to -1.5 eV in the calculated transmission functions, shaded region in Figure 5) in order to compare the computational results to our low-bias experimental conductance measurements. The ordering of the calculated transmission functions throughout this region is consistent with our experimental measurements. This agreement supports the importance of the limiting computational conformations considered here for understanding transport in these -SMe-linked molecules, and we conclude that the Au-S-C-C angle differences seen in Table 1 are relevant for the systems at hand.

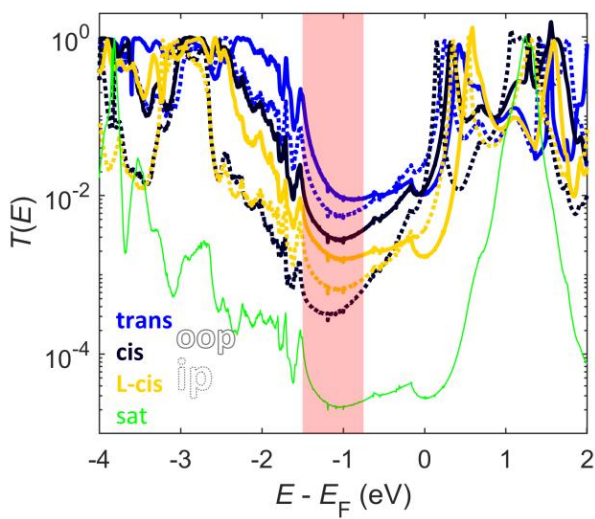

Figure 5. Overlaid transmission functions calculated for each of the conformations shown in Figure 4. For the trans, cis, and L-cis molecules, transmission in the oop and ip orientations are shown in solid and dotted lines, respectively. The red rectangle indicates the expected region of the experimental Fermi energy, $E_F$, (i.e., off-resonant but nearer the HOMO peak than the LUMO peak). The ordering of transmission functions within this region matches the experimentally measured conductances in Figure 2b.

We note that the calculated zero-bias transmission function of trans-OPV2-2SMe-ip is larger than that of cis-OPV2-2SMe-oop, despite the fact that the Au-S-C-C dihedrals are closer to 90º in the latter. This implies that while the linker orientation may be a large reason for the lower conductance of the cis molecules vis-à-vis trans-OPV2-2SMe, additional factors are also at play. We propose that the partially broken conjugation caused by ring twisting in cis-OPV2-2SMe, discussed above, explains this extra reduction in conductance. While L-cis-OPV2-2SMe overcomes this internal twisting problem, this may be counteracted by the locked-cis backbone keeping the Au-S-C-C angles even farther from the optimal value of 90º than in the unlocked cis structure (see Table 1). In addition, both the transmission functions in Figure 5 and gas-phase ionization energy calculations (SI section S.9) indicate that L-cis-OPV2-2SMe has a slightly larger transport gap than cis-OPV2-2SMe, which also partially counteracts the increase in conductance when untwisted. This provides another example of how complete structure-function relationships in single molecule quantum transport may require the inclusion of multiple competing factors, a central finding of this work.

To probe the impact of -SMe linker orientation on molecular conductance more systematically, we also performed NEGF-DFT calculations on a series of structures with both Au-S-C-C dihedral angles locked at values ranging from 0º to 90º. Rotation of these dihedrals can also be achieved in a more physically plausible way by varying the junction gap size, yielding qualitatively similar results (SI section S.10). We used the L-cis-OPV2-2SMe structure for this investigation to eliminate the confounding factor of ring-twisting degrees of freedom. The most striking feature of the transmission functions for this series of structures (Figure 6a) is that the valley between the HOMO- and LUMO-like transmission peaks becomes much deeper as both Au-S-C-C angles are rotated from 90° towards 0°. Examining the scattering states in the window where experimental transport is expected to occur reveals that in the 90° structure (Figure 6b), the electronic wavefunction flows efficiently from electrode to molecule, whereas in the 0° structure (Figure 6c) the wavefunction is almost completely reflected due to the fact that the orbital symmetry around the sulfur is near-perpendicular to the molecular π-system (Figure 6d). The transmission functions in Figure 6a thus provide direct evidence that the Au-S-C-C dihedral angles are important for controlling metal/molecule electronic coupling efficiency in OPV2-2SMe, in agreement with both the arguments made above and previous findings for conjugated molecules with -SMe linker groups.[31,33]

The transmission functions in Figure 6a reveal that rotating the Au-S-C-C dihedral angles from 90° towards 0° also has the effect of shifting the HOMO- and LUMO-like transmission peaks towards more negative energies. As shown in SI section S.11, qualitatively similar shifts occur for the molecular transport levels of these same structures in the gas phase, suggesting that this energy-level alignment effect is primarily caused by changes to the molecular structure on its own rather than changes to metal/molecule interactions. In particular, we find that changing the Au-S-C-C dihedral angle also changes the angle between the -SMe and its attached ring, which in turn controls how much the sulfur acts as an electron donor to the π-system. The fact that -SMe groups can act as molecular substituents in addition to their role as linker groups thus adds another layer of complexity that must be considered when employing -SMe and similar linkers for use in single-molecule transport applications.

By providing insight into how -SMe linker orientation affects both metal/molecule electronic coupling *and* energy-level alignment, the full set of transmission functions in Figure 6a thus significantly expands on previous studies, which connected -SMe orientation and conductance using simplified models[31] or only presented transmission functions for one or two -SMe orientations.[33,53]

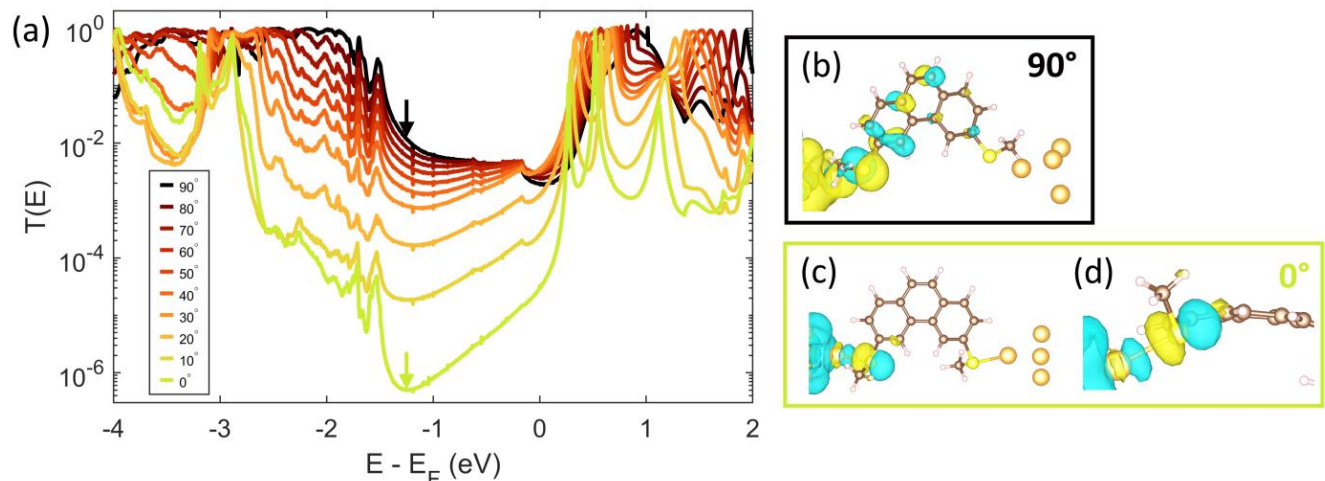

Figure 6. (a) Overlaid transmission functions calculated for L-cis-OPV2-2SMe with both Au-S-C-C dihedral angles locked at values ranging from 0° to 90°, demonstrating the pronounced effect this rotation has on molecular conductance. (b) Scattering state for the 90° transmission function from (a), calculated at $E - E_F$ = -1.25 eV (the dark red arrow in (a)). (c,d) Two different views of the scattering state for the 0° transmission function from (a), calculated at $E - E_F$ = -1.25 eV (the yellow arrow in (a)). In particular, the view in (d) shows how the wavefunction on the sulfur is misaligned with the π-system of the molecule, leading to poor electronic coupling efficiency and contributing to the suppressed transmission.

## 4. Conclusions

In summary, we show both experimentally and computationally the multifaceted reasons whereby single-molecule conductance is modified by the simple modification from cis- to trans-geometry in -SMe-linked stilbene derivatives. In particular, we find that a dominant role is played by the different ways in which cis and trans geometry constrain the binding conformations available to the -SMe linker groups, which in turn control the metal/molecule electronic coupling efficiency and even in-junction molecular electronic structure. Our study thus shows how two common structural attributes and design elements for single molecule electronics—cis vs. trans isomerization, and coupling efficiency of flexible linkers such as -SMe—can interact with each other in non-trivial ways to control molecular conductance. These lessons are relevant to single-molecule transport studies employing structural elements in molecular design that induce large geometric changes, and to those that rely on conformationally flexible linkers like -SMe as design elements. More generally, this reveals that, in many practical cases, "simple" single parameter structure-property rules and relationships may be insufficient, and instead more nuanced rules that take such interactions between effects into account are needed.

## ASSOCIATED CONTENT

**Supporting information:** This material is available free of charge *via* the Internet at http://pubs.acs.org.

Molecular Synthesis and Characterization; Datasets Used; Comment on cis-OPV2-1SMe Datasets; Use of Segment Clustering; Apparent Molecular Lengths; Preparation of Two-Probe Geometries for NEGT-DFT Calculations; NEGF-DFT Results for Single-Linker Molecules; Barrier to -SMe Rotation; Gas-Phase Ionization

Energies; Controlling Au-S Binding Angles Through Gap Size; and Electronic Structure Changes During -SMe Rotation.

## AUTHOR INFORMATION

### Corresponding Author

*Oliver L.A. Monti – Department of Chemistry and Biochemistry, University of Arizona, Tucson, Arizona 85721, United States; Department of Physics, University of Arizona, Tucson, Arizona 85721, United States; Email: monti@u.arizona.edu; Phone: ++ 520 626 1177; orcid.org/0000-0002-0974-7253.

### Authors

Nathan D. Bamberger – Department of Chemistry and Biochemistry, University of Arizona, Tucson, Arizona 85721, United States; orcid.org/0000-0001-5348-5695.

Dylan Dyer – Department of Chemistry and Biochemistry, University of Arizona, Tucson, Arizona 85721, United States; orcid.org/0000-0002-2747-9301.

Dawson Pursell – Department of Chemistry and Biochemistry, University of Arizona, Tucson, Arizona 85721, United States.

Keshaba N. Parida – Department of Chemistry and Biochemistry, University of Arizona, Tucson, Arizona 85721, United States; orcid.org/0000-0003-3454-0868.

Tarek H. El-Assaad – Department of Chemistry and Biochemistry, University of Arizona, Tucson, Arizona 85721, United States; orcid.org/0000-0002-7106-834X

Dominic V. McGrath – Department of Chemistry and Biochemistry, University of Arizona, Tucson, Arizona 85721, United States; orcid.org/0000-0001-9605-2224.

Manuel Smeu – Department of Physics, Binghamton University-SUNY, Binghamton, NY 13902, United States; orcid.org/0000-0001-9548-4623.

### Author Contributions

N.B., D.D., D.M., O.M, and M.S. conceived the research ideas. K.P. and T.E. synthesized the molecules, directed by D.M. N.B. and D.D. fabricated the MCBJ samples and collected the break junction data. M.S. performed the NEGF-DFT calculations and analysis. D.P. performed the gas-phase DFT calculations and analysis. N.B. wrote the manuscript with advice and input from all authors.

### Notes

The authors declare no competing financial interests.

## Acknowledgments

Financial support from the National Science Foundation, Award No. DMR-1708443, is gratefully acknowledged. Plasma etching of MCBJ samples was performed using a Plasmatherm reactive ion etcher acquired through an NSF MRI grant, Award No. ECCS-1725571. Segment clustering was performed using High Performance Computing (HPC) resources supported by the University of Arizona TRIF, UITS, and RDI and maintained by the UA Research Technologies department. Quality control was performed using a scanning electron microscope in the W. M. Keck Center for Nano-Scale Imaging in the Department of Chemistry and Biochemistry at the University of Arizona with funding from the W. M. Keck Foundation Grant. NEGF-DFT computations were performed on the Binghamton University HPC cluster, "Spiedie".

# References

(1) Heath, J. R. Molecular Electronics. *Annual Review of Materials Research* **2009**, *39*, 1–23. https://doi.org/10.1146/annurev-matsci-082908-145401.

(2) Forrest, S. R. The Path to Ubiquitous and Low-Cost Organic Electronic Appliances on Plastic. *Nature* **2004**, *428* (6986), 911–918. https://doi.org/10.1038/nature02498.

(3) Liu, Y.; Qiu, X.; Soni, S.; Chiechi, R. C. Charge Transport through Molecular Ensembles: Recent Progress in Molecular Electronics. *Chem. Phys. Rev.* **2021**, *2* (2), 021303. https://doi.org/10.1063/5.0050667.

(4) Ivie, J. A.; Bamberger, N. D.; Parida, K. N.; Shepard, S.; Dyer, D.; Saraiva-Souza, A.; Himmelhuber, R.; McGrath, D. V.; Smeu, M.; Monti, O. L. A. Correlated Energy-Level Alignment Effects Determine Substituent-Tuned Single-Molecule Conductance. *ACS Appl. Mater. Interfaces* **2021**, *13* (3), 4267–4277. https://doi.org/10.1021/acsami.0c19404.

(5) Su, T. A.; Neupane, M.; Steigerwald, M. L.; Venkataraman, L.; Nuckolls, C. Chemical Principles of Single-Molecule Electronics. *Nat Rev Mater* **2016**, *1* (3), 1–15. https://doi.org/10.1038/natrevmats.2016.2.

(6) Venkataraman, L.; Park, Y. S.; Whalley, A. C.; Nuckolls, C.; Hybertsen, M. S.; Steigerwald, M. L. Electronics and Chemistry: Varying Single-Molecule Junction Conductance Using Chemical Substituents. *Nano Lett.* **2007**, *7* (2), 502–506. https://doi.org/10.1021/nl062923j.

(7) Garner, M. H.; Li, H.; Chen, Y.; Su, T. A.; Shangguan, Z.; Paley, D. W.; Liu, T.; Ng, F.; Li, H.; Xiao, S.; Nuckolls, C.; Venkataraman, L.; Solomon, G. C. Comprehensive Suppression of Single-Molecule Conductance Using Destructive σ-Interference. *Nature* **2018**, *558* (7710), 415–419. https://doi.org/10.1038/s41586-018-0197-9.

(8) Kaliginedi, V.; Rudnev, A. V.; Moreno-García, P.; Baghernejad, M.; Huang, C.; Hong, W.; Wandlowski, T. Promising Anchoring Groups for Single-Molecule Conductance Measurements. *Phys. Chem. Chem. Phys.* **2014**, *16* (43), 23529–23539. https://doi.org/10.1039/C4CP03605K.

(9) Medina, S.; García-Arroyo, P.; Li, L.; Gunasekaran, S.; Stuyver, T.; José Mancheño, M.; Alonso, M.; Venkataraman, L.; L. Segura, J.; Cordón, J. C. Single-Molecule Conductance in a Unique Cross-Conjugated Tetra(Aminoaryl)-Ethene. *Chemical Communications* **2020**, *57* (5), 591–594. https://doi.org/10.1039/D0CC07124B.

(10) Aradhya, S. V.; Meisner, J. S.; Krikorian, M.; Ahn, S.; Parameswaran, R.; Steigerwald, M. L.; Nuckolls, C.; Venkataraman, L. Dissecting Contact Mechanics from Quantum Interference in Single-Molecule Junctions of Stilbene Derivatives. *Nano Lett.* **2012**, *12* (3), 1643–1647. https://doi.org/10.1021/nl2045815.

(11) Widawsky, J. R.; Darancet, P.; Neaton, J. B.; Venkataraman, L. Simultaneous Determination of Conductance and Thermopower of Single Molecule Junctions. *Nano Lett.* **2012**, *12* (1), 354–358. https://doi.org/10.1021/nl203634m.

(12) Kamenetska, M.; Quek, S. Y.; Whalley, A. C.; Steigerwald, M. L.; Choi, H. J.; Louie, S. G.; Nuckolls, C.; Hybertsen, M. S.; Neaton, J. B.; Venkataraman, L. Conductance and Geometry of Pyridine-Linked Single-Molecule Junctions. *J. Am. Chem. Soc.* **2010**, *132* (19), 6817–6821. https://doi.org/10.1021/ja1015348.

(13) Meisner, J. S.; Ahn, S.; Aradhya, S. V.; Krikorian, M.; Parameswaran, R.; Steigerwald, M.; Venkataraman, L.; Nuckolls, C. Importance of Direct Metal−π Coupling in Electronic Transport Through Conjugated Single-Molecule Junctions. *J. Am. Chem. Soc.* **2012**, *134* (50), 20440–20445. https://doi.org/10.1021/ja308626m.

(14) Meng, L.; Xin, N.; Hu, C.; Wang, J.; Gui, B.; Shi, J.; Wang, C.; Shen, C.; Zhang, G.; Guo, H.; Meng, S.; Guo, X. Side-Group Chemical Gating via Reversible Optical and Electric Control in a Single Molecule Transistor. *Nature Communications* **2019**, *10* (1), 1450. https://doi.org/10.1038/s41467-019-09120-1.

(15) Kim, Y. Photoswitching Molecular Junctions: Platforms and Electrical Properties. *ChemPhysChem* **2020**, *21* (21), 2368–2383. https://doi.org/10.1002/cphc.202000564.

(16) Martin, S.; Haiss, W.; Higgins, S. J.; Nichols, R. J. The Impact of E−Z Photo-Isomerization on Single Molecular Conductance. *Nano Lett.* **2010**, *10* (6), 2019–2023. https://doi.org/10.1021/nl9042455.

(17) Kim, Y.; Garcia-Lekue, A.; Sysoiev, D.; Frederiksen, T.; Groth, U.; Scheer, E. Charge Transport in Azobenzene-Based Single-Molecule Junctions. *Phys. Rev. Lett.* **2012**, *109* (22), 226801. https://doi.org/10.1103/PhysRevLett.109.226801.

(18) Zhang, C.; Du, M.-H.; Cheng, H.-P.; Zhang, X.-G.; Roitberg, A. E.; Krause, J. L. Coherent Electron Transport through an Azobenzene Molecule: A Light-Driven Molecular Switch. *Phys. Rev. Lett.* **2004**, *92* (15), 158301. https://doi.org/10.1103/PhysRevLett.92.158301.

(19) Osella, S.; Samorì, P.; Cornil, J. Photoswitching Azobenzene Derivatives in Single Molecule Junctions: A Theoretical Insight into the I/V Characteristics. *J. Phys. Chem. C* **2014**, *118* (32), 18721–18729. https://doi.org/10.1021/jp504582a.

(20) Zhang, C.; He, Y.; Cheng, H.-P.; Xue, Y.; Ratner, M. A.; Zhang, X.-G.; Krstic, P. Current-Voltage Characteristics through a Single Light-Sensitive Molecule. *Phys. Rev. B* **2006**, *73* (12), 125445. https://doi.org/10.1103/PhysRevB.73.125445.

(21) Zemanová Dieškova, M.; Štich, I.; Bokes, P. Rigidity of the Conductance of an Anchored Dithioazobenzene Optomechanical Switch. *Phys. Rev. B* **2013**, *87* (24), 245418. https://doi.org/10.1103/PhysRevB.87.245418.

(22) Mativetsky, J. M.; Pace, G.; Elbing, M.; Rampi, M. A.; Mayor, M.; Samorì, P. Azobenzenes as Light-Controlled Molecular Electronic Switches in Nanoscale Metal−Molecule−Metal Junctions. *J. Am. Chem. Soc.* **2008**, *130* (29), 9192–9193. https://doi.org/10.1021/ja8018093.

(23) Lin, L.; Jiang, J.; Luo, Y. Elastic and Inelastic Electron Transport in Metal–Molecule(s)–Metal Junctions. *Physica E: Low-dimensional Systems and Nanostructures* **2013**, *47*, 167–187. https://doi.org/10.1016/j.physe.2012.10.017.

(24) Vonlanthen, D.; Mishchenko, A.; Elbing, M.; Neuburger, M.; Wandlowski, T.; Mayor, M. Chemically Controlled Conductivity: Torsion-Angle Dependence in a Single-Molecule Biphenyldithiol Junction. *Angewandte Chemie International Edition* **2009**, *48* (47), 8886–8890. https://doi.org/10.1002/anie.200903946.

(25) Venkataraman, L.; Klare, J. E.; Nuckolls, C.; Hybertsen, M. S.; Steigerwald, M. L. Dependence of Single-Molecule Junction Conductance on Molecular Conformation. *Nature* **2006**, *442* (7105), 904–907. https://doi.org/10.1038/nature05037.

(26) Mishchenko, A.; Zotti, L. A.; Vonlanthen, D.; Bürkle, M.; Pauly, F.; Cuevas, J. C.; Mayor, M.; Wandlowski, T. Single-Molecule Junctions Based on Nitrile-Terminated Biphenyls: A Promising New Anchoring Group. *J. Am. Chem. Soc.* **2011**, *133* (2), 184–187. https://doi.org/10.1021/ja107340t.

(27) Li, X.; Tan, Z.; Huang, X.; Bai, J.; Liu, J.; Hong, W. Experimental Investigation of Quantum Interference in Charge Transport through Molecular Architectures. *J. Mater. Chem. C* **2019**, *7* (41), 12790–12808. https://doi.org/10.1039/C9TC02626F.

(28) Park, Y. S.; Whalley, A. C.; Kamenetska, M.; Steigerwald, M. L.; Hybertsen, M. S.; Nuckolls, C.; Venkataraman, L. Contact Chemistry and Single-Molecule Conductance: A Comparison of
16

Phosphines, Methyl Sulfides, and Amines. *J. Am. Chem. Soc.* **2007**, *129* (51), 15768–15769. https://doi.org/10.1021/ja0773857.
(29) Dell, E. J.; Capozzi, B.; Xia, J.; Venkataraman, L.; Campos, L. M. Molecular Length Dictates the Nature of Charge Carriers in Single-Molecule Junctions of Oxidized Oligothiophenes. *Nature Chemistry* **2015**, *7* (3), 209–214. https://doi.org/10.1038/nchem.2160.
(30) Jiang, F.; Trupp, D.; Algethami, N.; Zheng, H.; He, W.; Alqorashi, A.; Zhu, C.; Tang, C.; Li, R.; Liu, J.; Sadeghi, H.; Shi, J.; Davidson, R.; Korb, M.; Naher, M.; Sobolev, A. N.; Sangtarash, S.; Low, P. J.; Hong, W.; Lambert, C. Turning the Tap: Conformational Control of Quantum Interference to Modulate Single Molecule Conductance. *Angewandte Chemie* **2019**, *131* (52), 19163–19169. https://doi.org/10.1002/ange.201909461.
(31) Park, Y. S.; Widawsky, J. R.; Kamenetska, M.; Steigerwald, M. L.; Hybertsen, M. S.; Nuckolls, C.; Venkataraman, L. Frustrated Rotations in Single-Molecule Junctions. *J. Am. Chem. Soc.* **2009**, *131* (31), 10820–10821. https://doi.org/10.1021/ja903731m.
(32) Sagan, C.; Jiang, Y.; Caban, F.; Snaider, J.; Amell, R.; Wei, S.; Florio, G. M. Oligofluorene Molecular Wires: Synthesis and Single-Molecule Conductance. *J. Phys. Chem. C* **2017**, *121* (45), 24945–24953. https://doi.org/10.1021/acs.jpcc.7b07713.
(33) Wang, M.; Wang, Y.; Sanvito, S.; Hou, S. The Low-Bias Conducting Mechanism of Single-Molecule Junctions Constructed with Methylsulfide Linker Groups and Gold Electrodes. *J. Chem. Phys.* **2017**, *147* (5), 054702. https://doi.org/10.1063/1.4996745.
(34) Bamberger, N. D.; Ivie, J. A.; Parida, K.; McGrath, D. V.; Monti, O. L. A. Unsupervised Segmentation-Based Machine Learning as an Advanced Analysis Tool for Single Molecule Break Junction Data. *J. Phys. Chem. C* **2020**, *124* (33), 18302–18315. https://doi.org/10.1021/acs.jpcc.0c03612.
(35) Bamberger, N. D.; Dyer, D.; Parida, K. N.; McGrath, D. V.; Monti, O. L. A. Grid-Based Correlation Analysis to Identify Rare Quantum Transport Behaviors. *J. Phys. Chem. C* **2021**, *125* (33), 18297–18307. https://doi.org/10.1021/acs.jpcc.1c04794.
(36) Johnson, T. K.; Ivie, J. A.; Jaruvang, J.; Monti, O. L. A. Fast Sensitive Amplifier for Two-Probe Conductance Measurements in Single Molecule Break Junctions. *Review of Scientific Instruments* **2017**, *88* (3), 033904. https://doi.org/10.1063/1.4978962.
(37) Ohnishi, H.; Kondo, Y.; Takayanagi, K. Quantized Conductance through Individual Rows of Suspended Gold Atoms. *Nature* **1998**, *395* (6704), 780–783. https://doi.org/10.1038/27399.
(38) Cabosart, D.; El Abbassi, M.; Stefani, D.; Frisenda, R.; Calame, M.; van der Zant, H. S. J.; Perrin, M. L. A Reference-Free Clustering Method for the Analysis of Molecular Break-Junction Measurements. *Appl. Phys. Lett.* **2019**, *114* (14), 143102. https://doi.org/10.1063/1.5089198.
(39) Neese, F. The ORCA Program System. *WIREs Computational Molecular Science* **2012**, *2* (1), 73–78. https://doi.org/10.1002/wcms.81.
(40) Becke, A. D. Density-functional Thermochemistry. III. The Role of Exact Exchange. *J. Chem. Phys.* **1993**, *98* (7), 5648–5652. https://doi.org/10.1063/1.464913.
(41) Lee, C.; Yang, W.; Parr, R. G. Development of the Colle-Salvetti Correlation-Energy Formula into a Functional of the Electron Density. *Phys. Rev. B* **1988**, *37* (2), 785–789. https://doi.org/10.1103/PhysRevB.37.785.
(42) Kresse, G.; Hafner, J. Ab Initio Molecular Dynamics for Liquid Metals. *Phys. Rev. B* **1993**, *47* (1), 558–561. https://doi.org/10.1103/PhysRevB.47.558.
(43) Kresse, G.; Furthmüller, J. Efficient Iterative Schemes for Ab Initio Total-Energy Calculations Using a Plane-Wave Basis Set. *Phys. Rev. B* **1996**, *54* (16), 11169–11186. https://doi.org/10.1103/PhysRevB.54.11169.

(44) Perdew, J. P.; Burke, K.; Ernzerhof, M. Generalized Gradient Approximation Made Simple. *Phys. Rev. Lett.* **1996**, *77* (18), 3865–3868. https://doi.org/10.1103/PhysRevLett.77.3865.

(45) Grimme, S.; Antony, J.; Ehrlich, S.; Krieg, H. A Consistent and Accurate Ab Initio Parametrization of Density Functional Dispersion Correction (DFT-D) for the 94 Elements H-Pu. *J. Chem. Phys.* **2010**, *132* (15), 154104. https://doi.org/10.1063/1.3382344.

(46) Taylor, J.; Guo, H.; Wang, J. Ab Initio Modeling of Quantum Transport Properties of Molecular Electronic Devices. *Phys. Rev. B* **2001**, *63* (24), 245407. https://doi.org/10.1103/PhysRevB.63.245407.

(47) Waldron, D.; Haney, P.; Larade, B.; MacDonald, A.; Guo, H. Nonlinear Spin Current and Magnetoresistance of Molecular Tunnel Junctions. *Phys. Rev. Lett.* **2006**, *96* (16), 166804. https://doi.org/10.1103/PhysRevLett.96.166804.

(48) Smeu, M.; Monti, O. L. A.; McGrath, D. Phenalenyls as Tunable Excellent Molecular Conductors and Switchable Spin Filters. *Phys. Chem. Chem. Phys.* **2021**, *23* (42), 24106–24110. https://doi.org/10.1039/D1CP04037E.

(49) Metzger, R. M. Unimolecular Electronics. *Chem. Rev.* **2015**, *115* (11), 5056–5115. https://doi.org/10.1021/cr500459d.

(50) Martín, S.; Grace, I.; Bryce, M. R.; Wang, C.; Jitchati, R.; Batsanov, A. S.; Higgins, S. J.; Lambert, C. J.; Nichols, R. J. Identifying Diversity in Nanoscale Electrical Break Junctions. *J. Am. Chem. Soc.* **2010**, *132* (26), 9157–9164. https://doi.org/10.1021/ja103327f.

(51) Yoshida, K.; Pobelov, I. V.; Manrique, D. Z.; Pope, T.; Mészáros, G.; Gulcur, M.; Bryce, M. R.; Lambert, C. J.; Wandlowski, T. Correlation of Breaking Forces, Conductances and Geometries of Molecular Junctions. *Sci Rep* **2015**, *5*, 9002. https://doi.org/10.1038/srep09002.

(52) Wu, S.; González, M. T.; Huber, R.; Grunder, S.; Mayor, M.; Schönenberger, C.; Calame, M. Molecular Junctions Based on Aromatic Coupling. *Nat Nano* **2008**, *3* (9), 569–574. https://doi.org/10.1038/nnano.2008.237.

(53) Li, H.; Garner, M. H.; Shangguan, Z.; Zheng, Q.; Su, T. A.; Neupane, M.; Li, P.; Velian, A.; Steigerwald, M. L.; Xiao, S.; Nuckolls, C.; Solomon, G. C.; Venkataraman, L. Conformations of Cyclopentasilane Stereoisomers Control Molecular Junction Conductance. *Chem. Sci.* **2016**, *7* (9), 5657–5662. https://doi.org/10.1039/C6SC01360K.

(54) Batra, A.; Darancet, P.; Chen, Q.; Meisner, J. S.; Widawsky, J. R.; Neaton, J. B.; Nuckolls, C.; Venkataraman, L. Tuning Rectification in Single-Molecular Diodes. *Nano Lett.* **2013**, *13* (12), 6233–6237. https://doi.org/10.1021/nl403698m.

(55) Verzijl, C. J. O.; Celis Gil, J. A.; Perrin, M. L.; Dulić, D.; van der Zant, H. S. J.; Thijssen, J. M. Image Effects in Transport at Metal-Molecule Interfaces. *J. Chem. Phys.* **2015**, *143* (17), 174106. https://doi.org/10.1063/1.4934882.

(56) Toher, C.; Filippetti, A.; Sanvito, S.; Burke, K. Self-Interaction Errors in Density-Functional Calculations of Electronic Transport. *Phys. Rev. Lett.* **2005**, *95* (14), 146402. https://doi.org/10.1103/PhysRevLett.95.146402.

(57) Sangtarash, S.; Huang, C.; Sadeghi, H.; Sorohhov, G.; Hauser, J.; Wandlowski, T.; Hong, W.; Decurtins, S.; Liu, S.-X.; Lambert, C. J. Searching the Hearts of Graphene-like Molecules for Simplicity, Sensitivity, and Logic. *J. Am. Chem. Soc.* **2015**, *137* (35), 11425–11431. https://doi.org/10.1021/jacs.5b06558.

# Supporting Information for:
# Beyond Simple Structure-Function Relationships: Interplay Between Cis/Trans Isomerization and Geometrically Constrained Metal/Molecule Coupling Efficiency in Single-Molecule Junctions

Nathan D. Bamberger[1], Dylan Dyer[1], Keshaba N. Parida[1], Tarek H. El Assaad,[1] Dawson Pursell[1], Dominic V. McGrath[1], Manuel Smeu[2] and Oliver L.A. Monti[1,3,*]

[1]Department of Chemistry and Biochemistry, University of Arizona, 1306 E. University Blvd., Tucson, Arizona 85721, USA

[2]Department of Physics, Binghamton University – SUNY, 4400 Vestal Parkway East, Binghamton, New York, 13902, USA

[3]Department of Physics, University of Arizona, 1118 E. Fourth Street, Tucson, Arizona 85721, USA

## Contents

## S.1 Molecular Synthesis and Characterization

### S.1.1 Synthetic Procedures

All reactions were carried out using oven-dried (120 °C) glassware. All commercially available reagents and common solvents were used without further purification. Reaction mixtures were magnetically stirred. Reaction progress was monitored by thin layer chromatography using Merck Silica Gel 60 F254 plates which were visualized by fluorescence quenching under UV light. Photocyclization was performed in an RPR-100 Rayonet PhotoChemical Reactor© using RPR-3500Å lamps as light source. Flash chromatographic purification was performed on silica 32-63,

60 Å under positive nitrogen pressure. Concentration under reduced pressure was performed by rotary evaporation at <40 °C at the appropriate pressure. Structural characterization data was obtained on commercially available instrumentation with specifications as indicated in the experimental descriptions.

*Synthesis of (Z)-1,2-bis(4-(methylthio)phenyl)ethene (**cis-OPV2-2SMe**):*

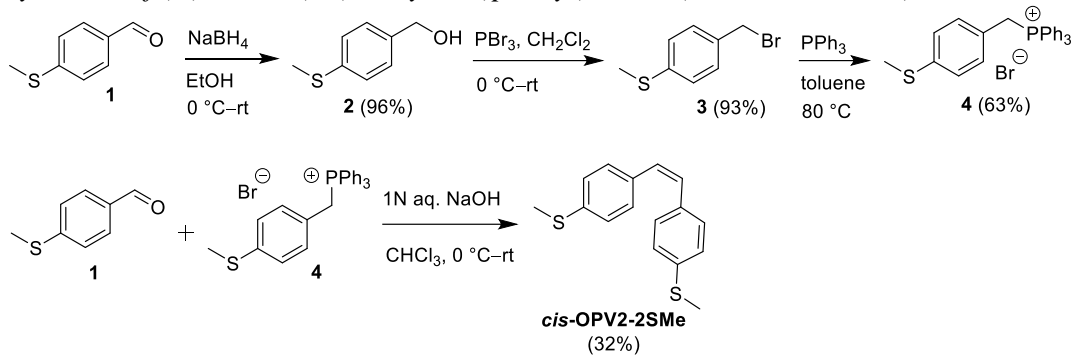

**(4-(Methylthio)phenyl)methanol (2):**[1] The reduction was performed using the literature report with modification.[1] To a stirred, cold (0 °C) solution of 4-(methylthio)benzaldehyde (**1**) (5.00 g, 32.9 mmol) in EtOH (30 mL) was added NaBH$_4$ (0.62 g, 16.42 mmol) portion-wise and stirring was continued at rt for overnight. The reaction was quenched at 0 °C with slow addition of water (30 mL). Solvent was removed under vacuum, and the residue was extracted with CH$_2$Cl$_2$ (3x30 mL). The combined organic layer was dried (MgSO$_4$), filtered, and concentrated and the residue subjected to column chromatographic purification to obtained alcohol **2** (4.89 g, 96%) as a colorless solid. **R$_f$**: 0.3 (1:3 *v/v* EtOAc/hexane); **$^1$H NMR** (400 MHz, CDCl$_3$): δ 7.24–7.30 (m, 4H), 4.65 (d, *J* = 5.7 Hz, 2H), 2.49 (s, 3H), 1.68 (t, *J* = 5.7 Hz, 1H) ppm.

**(4-(Bromomethyl)phenyl)(methyl)sulfane (3):**[2] Bromination was performed using a literature report with slight modification.[2] To a stirred, cold (0 °C) solution of (4-(methylthio)phenyl)methanol (**2**) (2.0 g, 13.0 mmol) in CH$_2$Cl$_2$ (20 mL) was slowly added PBr$_3$ (0.74 mL, 7.78 mmol) and stirring was continued at rt overnight. The reaction was quenched at 0 °C with slow addition of water (30 mL) followed by brine (10 mL). The mixture was extracted with CH$_2$Cl$_2$ (2x10 mL) and the combined organic layer was washed with aqueous saturated NaHCO$_3$ (10 mL). The organic layer was dried (MgSO$_4$), filtered and concentrated to provide **3** (2.61 g, 93%) as colorless solid, which was taken on without further purification. **R$_f$**: 0.5 (1:19 *v/v* EtOAc/hexane); **$^1$H NMR** (400 MHz, CDCl$_3$): δ 7.31 (d, *J* = 8.35 Hz, 2H), 7.21 (d, *J* = 8.35 Hz, 2H), 4.48 (s, 2H), 2.49 (s, 3H) ppm.

**(4-(Methylthio)benzyl)triphenylphosphonium bromide (4):**[3] The reaction was performed using a literature report with modification.[4] A mixture of (4-(bromomethyl)phenyl)(methyl)sulfane (**3**) (0.5 g, 2.3 mmol), PPh$_3$ (0.91 g, 3.45 mmol), and toluene (10 mL) was heated at 80 °C under N$_2$ for 6 h. The reaction mixture was allowed to cool to rt after which time was added CH$_2$Cl$_2$ (30 mL) and hexane (30 mL). The resulting slurry was filtered and the residue was air dried to afford **4** (0.7 g, 63%) as a white solid, which was taken on without further purification.

**(Z)-1,2-Bis(4-(methylthio)phenyl)ethene** (*cis*-**OPV2-2SMe**): The Wittig coupling was
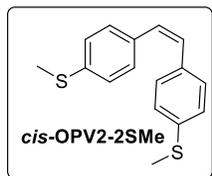
performed using a literature report with modification.[4] To a solution of aldehyde **1** (0.19 g, 1.23 mmol), phosphonium salt **4** (0.65 g, 1.35 mmol), and CHCl$_3$ (15 mL) was added aq. 1N NaOH (10 mL) at 0 °C. The reaction mixture was stirred at rt overnight. The reaction was quenched with H$_2$O (15 mL) and the resulting mixture was extracted with CH$_2$Cl$_2$ (3x10 mL). The combined organic layer was dried (MgSO$_4$), filtered, and concentrated, and the residue was subjected to column chromatographic purification to obtain desired ***cis*-OPV2-2SMe** (106 mg, 32%) as a colorless solid in addition a mixture of *E* and *Z* isomers (210 mg, 63%). **R$_f$**: 0.5 (1:19 *v/v* EtOAc/hexane); **$^1$H NMR** (400 MHz, CDCl$_3$): δ 7.19 (d, *J* = 8.5 Hz, 4H), 7.11 (d, *J* = 8.5 Hz, 4H), 6.5 (s, 2H), 2.48 (s, 6H) ppm; **$^{13}$C NMR** (101 MHz, CDCl$_3$): δ 137.4, 134.1, 129.5, 129.4, 126.2, 15.7 ppm, **LRMS** (MS+): m/z calculated for C$_{16}$H$_{17}$S$_2$ 273.0773 [M+H]$^+$; found 273.15.

**3,6-Bis(methylthio)phenanthrene** (**L-*cis*-OPV2-2SMe**): The photo-cyclization was performed
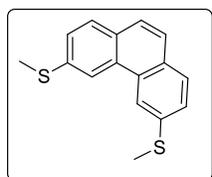
using literature report with modification.[5] A solution of (*E*)-1,2-bis(4-(methylthio)phenyl)ethene[6] (0.075 g, 0.275 mmol) and iodine (0.038 g, 0.151 mmol) in toluene (0.5 L) was irradiated (350 nm) under open-air condition at rt for 24 h. After the irradtiation, saturated aqueous Na$_2$S$_2$O$_3$ (10 mL) was added and the mixture was extracted with ethyl acetate (3x10 mL). The combined organic layer was dried (MgSO$_4$), filtered, concentrated, and the residue was subjected to column chromatographic purification to obtain desired **L-*cis*-OPV2-2SMe** (28 mg, 37%) as a colorless solid. **R$_f$**: 0.55 (1:19 *v/v* EtOAc/hexane); **$^1$H NMR** (400 MHz, CDCl$_3$): δ 8.46 (d, *J* = 1.8 Hz, 2H), 7.79 (d, *J* = 8.3 Hz, 2H), 7.62 (s, 2H), 7.52 (dd, *J* = 8.4, 1.8 Hz, 2H), 2.67 (s, 6H) ppm. **$^{13}$C NMR** (101 MHz, CDCl$_3$): δ 137.0, 130.3, 130.0, 129.1, 126.3, 126.1, 120.2, 16.6 ppm; **LRMS** (MS+): m/z calculated for C$_{16}$H$_{15}$S$_2$ 271.0615 [M+H]$^+$; found 271.12.

**1,2-Bis(4-(methylthio)phenyl)ethane** (**sat-OPV2-2SMe**):[7] To a slurry of Mg (0.134 g, 5.526
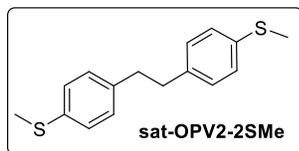
mmol), I$_2$ (0.116 g, 0.46 mmol), and Et$_2$O (30 mL), was added a solution of **3** (1.0 g, 4.6 mmol, in 5 mL Et$_2$O) was added to it. The reaction mixture was stirred at rt for 1 h, followed by addition of HCO$_2$Me (0.138 g, 2.3 mmol, in 5 mL Et$_2$O) and stirring overnight. The reaction was quenched at 0 °C with H$_2$O and the mixture was extracted with Et$_2$O (2x20 mL). The combined organic layer was dried (MgSO$_4$), filtered, and evaporated, and the residue was subjected to column chromatographic purification to gave compound **sat-OPV2-2SMe** (216 mg, 34%) as a colorless solid. **R$_f$**: 0.7 (1:19 *v/v* EtOAc/hexane); **$^1$H NMR** (400 MHz, CDCl$_3$): δ 7.15–7.23 (m, 4H), 7.04–7.12 (m, 4H), 2.86 (s, 4H), 2.48 (s, 6H) ppm. **$^{13}$C NMR** (101 MHz, CDCl$_3$): δ 138.8, 135.6, 129.2, 127.2, 37.4, 16.4 ppm; **LRMS** (MS+): m/z calculated for C$_{16}$H$_{18}$S$_2$ 275.0928 [M+H]$^+$; found 275.17.

*Synthesis of (E)/(Z)-Methyl(4-styrylphenyl)sulfane (**trans-/cis-OPV2-1SMe**):*

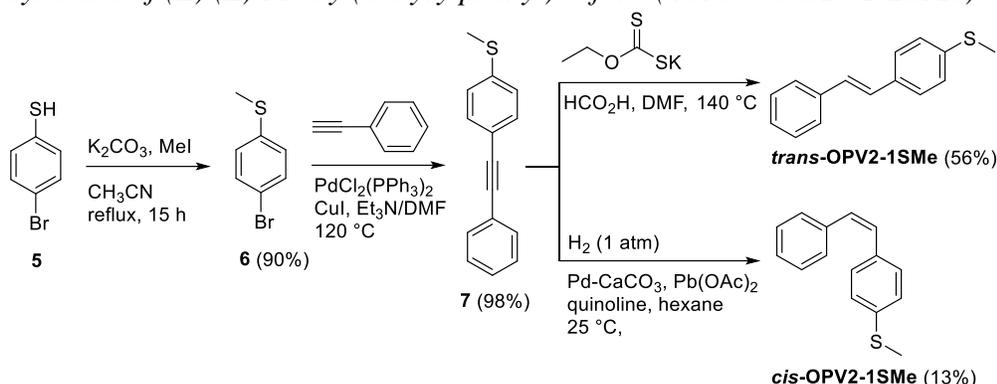

**(4-Bromophenyl)(methyl)sulfane (6):**[8] A mixture of MeCN (30 mL), 4-bromobenzenethiol (**5**)

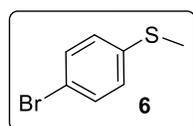

(3.50 g, 18.5 mmol), and $K_2CO_3$ (5.11 g, 37.0 mmol) was stirred for 5 min at rt. MeI (5.25 g, 37.0 mmol) was added and the mixture was then maintained at reflux for 15 h. The reaction mixture was cooled to rt, poured into water (300 mL), and the mixture was extracted with $CH_2Cl_2$ (20 mL*5). The combined organic layer was washed with 10% $K_2CO_3$ (3x30 mL) and brine (3x30 mL). The resulting solution was dried ($MgSO_4$), filtered, and concentrated under reduced pressure. The residue was recrystallized from hexanes at -78 °C. Compound **6** (3.40 g, 90%) was obtained as colorless crystals. $R_f$: 0.43 (hexane). **$^1$H NMR** (500 MHz, $CDCl_3$): δ 7.40 (d, $J$ = 8.6 Hz, 2H), 7.12 (d, $J$ = 8.6 Hz, 2H), 2.46 (s, 3H) ppm.

**Methyl(4-(phenylethynyl)phenyl)sulfane (7):**[9] In a 200 mL oven-dried three-neck round-bottomed flask equipped with a magnetic stirrer, freshly-prepared $PdCl_2(PPh_3)_2$ (175 mg, 0.25 mmol),[10] purified CuI (95 mg, 0.50 mmol),[11] and (4-bromophenyl)(methyl)sulfane (**6**) (1.01 g, 5.0 mmol) were mixed. The flask was degassed and backfilled with argon (5 min*3). In two separate flasks, dry $Et_3N$ (25 mL) and a solution of phenylacetylene (0.77 g, 7.5 mmol) in dry DMF (25 mL) were purged via bubbling with nitrogen for 25 minutes each and were transferred by cannula into the three-neck flask sequentially in that order. The three-neck flask was degassed and backfilled with argon (5 min), and stirred at 120 °C in the dark for 15 h. The reaction mixture was cooled to rt and quenched by simultaneous addition of 1M $NH_4Cl$ (25 mL) and $CH_2Cl_2$ (25 mL). The mixture was extracted with $CH_2Cl_2$ (3x25 mL), and the combined organic layer was washed with brine (3x25 mL), dried ($MgSO_4$), filtered, and concentrated under reduced pressure. The residue was subjected to chromatographic purification followed by recrystallization from MeOH to isolate alkyne **7** (1.10 g, 98%) as colorless crystals. $R_f$: 0.38 (2:98 v/v EtOAc/hexanes). **$^1$H NMR** (500 MHz, $CDCl_3$): δ 7.53 (d, $J$ = 7.9 Hz, 2H), 7.45 (d, $J$ = 8.6 Hz, 2H), 7.32–7.35 (m, 3H), 7.22 (d, $J$ = 8.6 Hz, 2H), 2.50 (s, 3H) ppm.

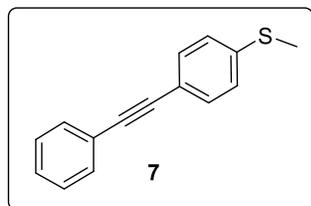

**(E)-Methyl(4-styrylphenyl)sulfane (trans-OPV2-1SMe):**[12,13] The reduction was carried according to a literature report.[14] In a pressure vessel tube, a mixture of methyl(4-(phenylethynyl)phenyl)sulfane (**7**) (112 mg, 0.50 mmol), potassium ethylxanthate (160 mg, 1.0 mmol), $HCO_2H$ (46 mg, 1.0 mmol), deionized $H_2O$ (90 mg, 5.0 mmol), and DMF (2 mL) was stirred at 140 °C for 4 h. The reaction was cooled, pressure was released, and the mixture was poured into water (20 mL). The resulting mixture was extracted with CH$_2$Cl$_2$ (3x20 mL) and the combined organic layer was washed with brine (3x20 mL), dried (MgSO$_4$), filtered, and concentrated. The residue was subjected to column chromatographic purification to obtain the desired ***trans*-OPV2-1SMe** (63 mg, 56%) as a colorless solid. **R**$_f$: 0.20 (1:99 *v/v* EtOAc/hexanes); **$^1$H NMR** (500 MHz, CDCl$_3$): δ 7.51 (d, *J* = 7.6 Hz, 2H), 7.45 (d, *J* = 8.4 Hz, 2H), 7.37 (t, *J* = 7.6 Hz, 2H), 7.23–7.27 (m, 3H), 7.06 (s, 2H), 2.51 (s, 3H) ppm; **$^{13}$C NMR** (126 MHz, CDCl$_3$) δ 137.8, 137.3, 134.3, 128.7, 128.1, 128.0, 127.55, 126.9, 126.7, 126.4, 15.8.

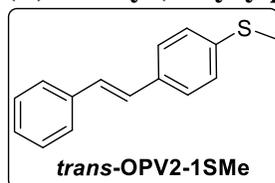

**(Z)-Methyl(4-styrylphenyl)sulfane** (***cis*-OPV2-1SMe**):[15] The Lindlar reduction method was adapted from the literature with modification.[16] In a 10 mL round-bottomed flask, a slurry of methyl(4-(phenylethynyl)phenyl)sulfane (**7**) (112 mg, 0.50 mmol) Lindlar's catalyst (12 mg), and hexane (5 mL) was degassed and backfilled with hydrogen gas (5 min*3). The resulting mixture was stirred at rt under hydrogen atmosphere for 15 h. The reaction mixture was filtered through celite and concentrated, and the residue was subjected to column chromatographic purification to isolate ***cis*-OPV2-1SMe** (15 mg, 13%) as colorless oil. **R**$_f$: 0.20 (1:99 *v/v* EtOAc/hexane). **$^1$H NMR** (500 MHz, CDCl$_3$): δ 7.20–7.27 (m, 5H), 7.18 (d, *J* = 8.4 Hz, 2H), 7.10 (d, *J* = 8.4 Hz, 2H), 6.59 (d, *J* = 12.2 Hz, 1H), 6.54 (d, *J* = 12.2 Hz, 1H), 2.46 (s, 3H) ppm. **$^{13}$C NMR** (126 MHz, CDCl$_3$): δ 137.3, 137.2, 133.9, 130.0, 129.6, 129.3, 128.8, 128.3, 127.1, 126.1, 15.6 ppm.

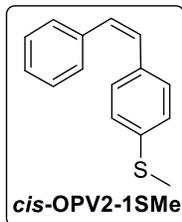
*cis*-OPV2-1SMe

### S.1.2 NMR Characterization

**Nuclear magnetic resonance** (NMR) spectra were recorded on Bruker Avance III 400 or Bruker NEO 500 spectrometer operating at 400/500 MHz ($^1$H) and 101/126 MHz ($^{13}$C), respectively. Chemical shifts (δ) are reported in parts per million (ppm) and referenced based on the NMR solvent resonance for both $^1$H and $^{13}$C. TMS and residual NMR solvent peaks are not labelled. All $^{13}$C spectra were measured with complete proton decoupling. Data are reported as follows: s = singlet, d = doublet, t = triplet, m = multiplet, coupling constants (*J*) in Hz. MNova software was used to plot NMR spectra.

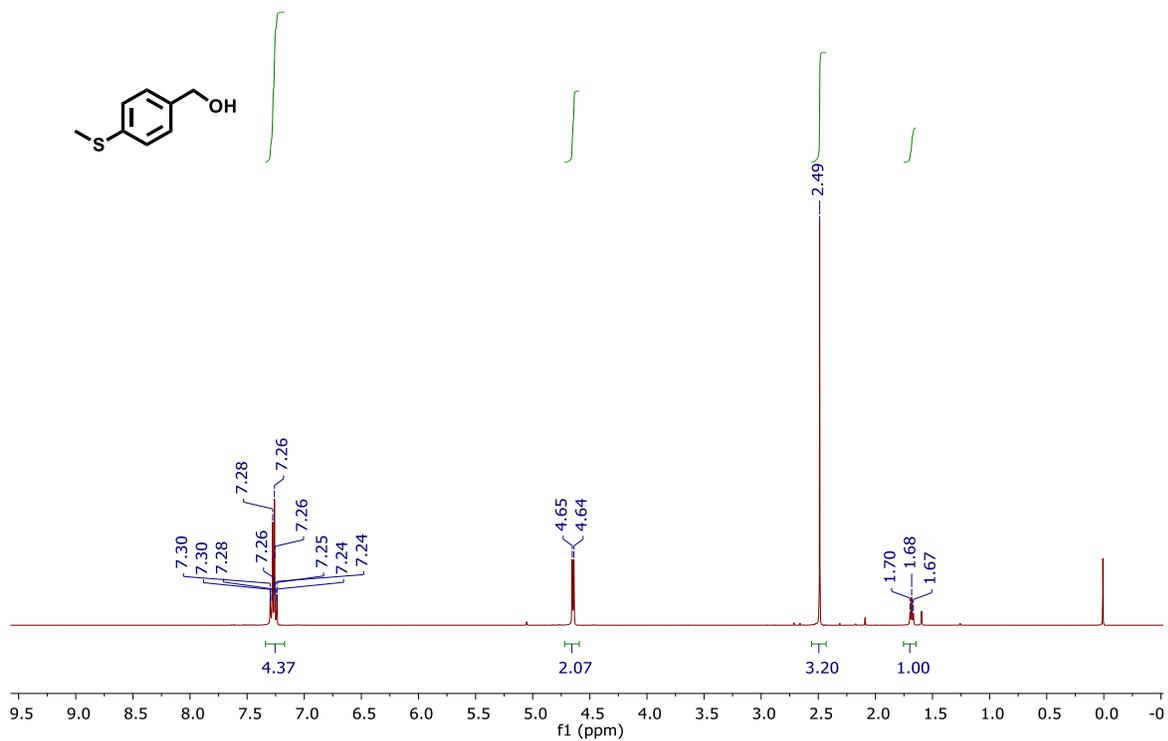

Figure S1. ¹H NMR (400 MHz) spectrum of (4-(methylthio)phenyl)methanol (**2**) in CDCl₃.

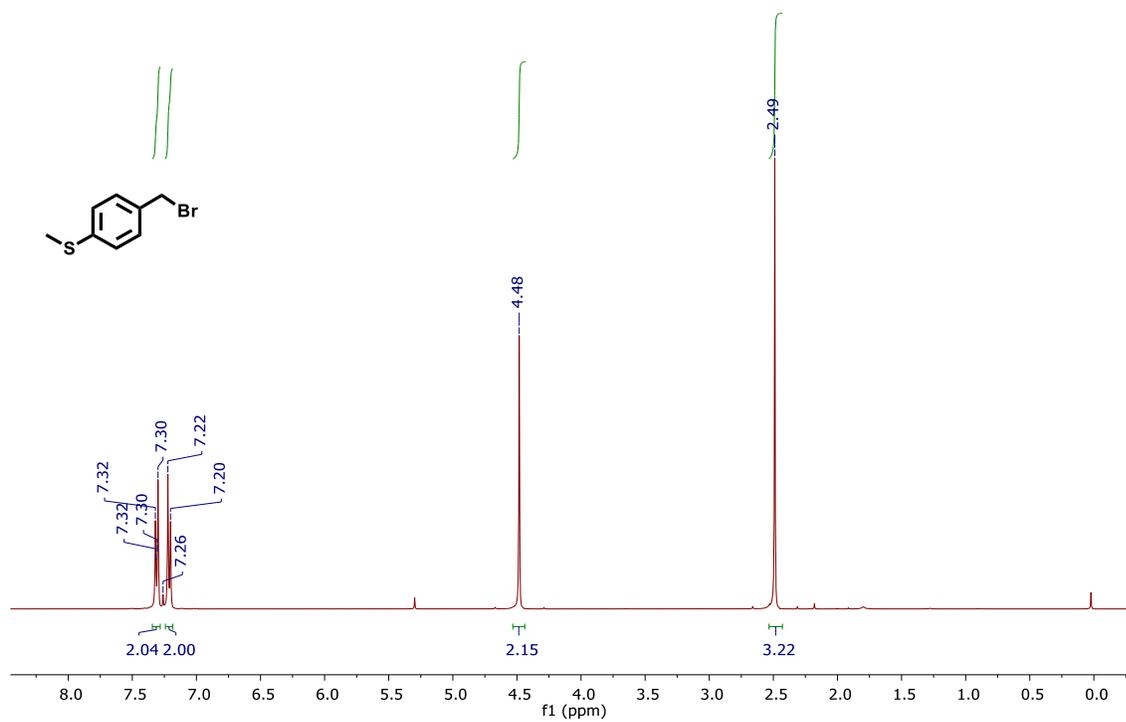

Figure S2. ¹H NMR (400 MHz) spectrum of (4-(bromomethyl)phenyl)(methyl)sulfane (**3**) in CDCl₃.

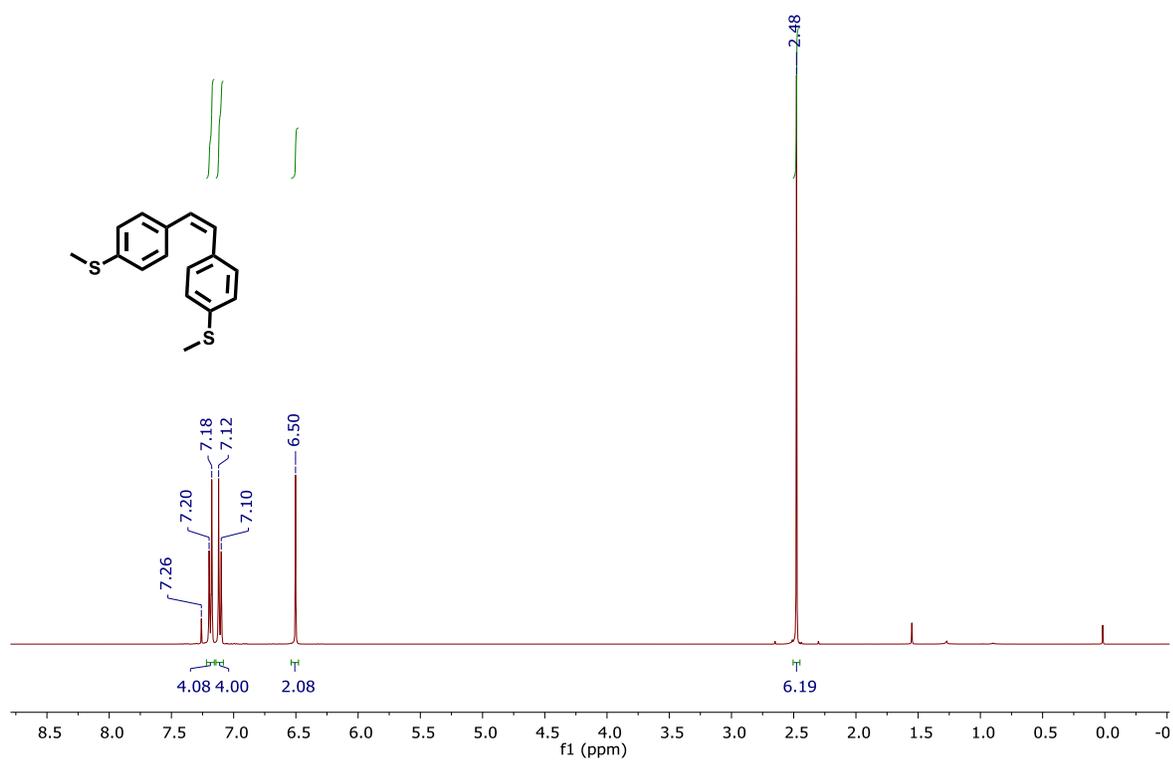

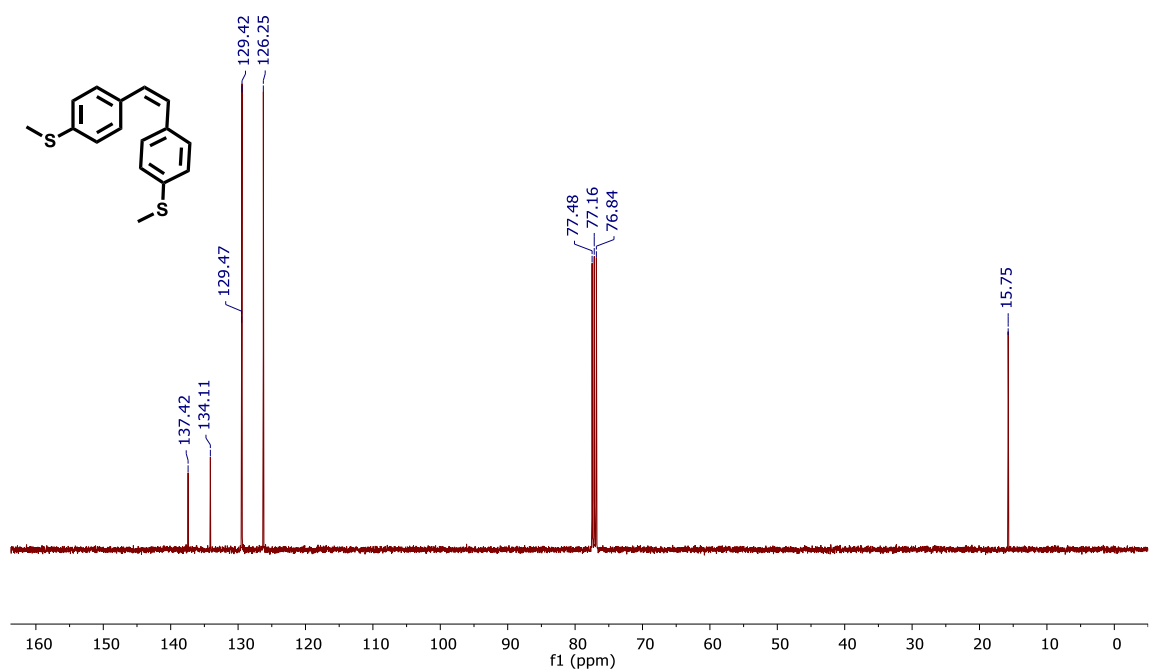

Figure S3. $^1$H NMR (400 MHz) and $^{13}$C NMR (101 MHz) spectra of (Z)-1,2-bis(4-(methylthio)phenyl)ethene (*cis*-**OPV2-2SMe**) in CDCl$_3$.

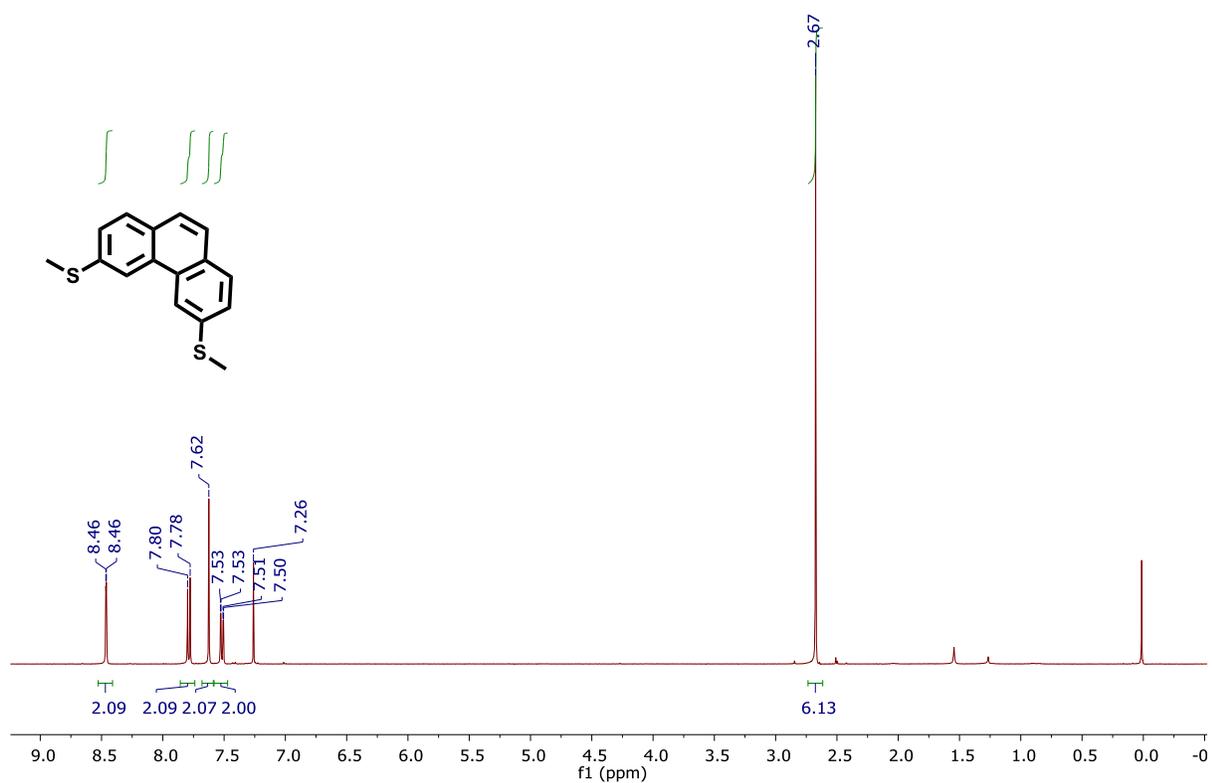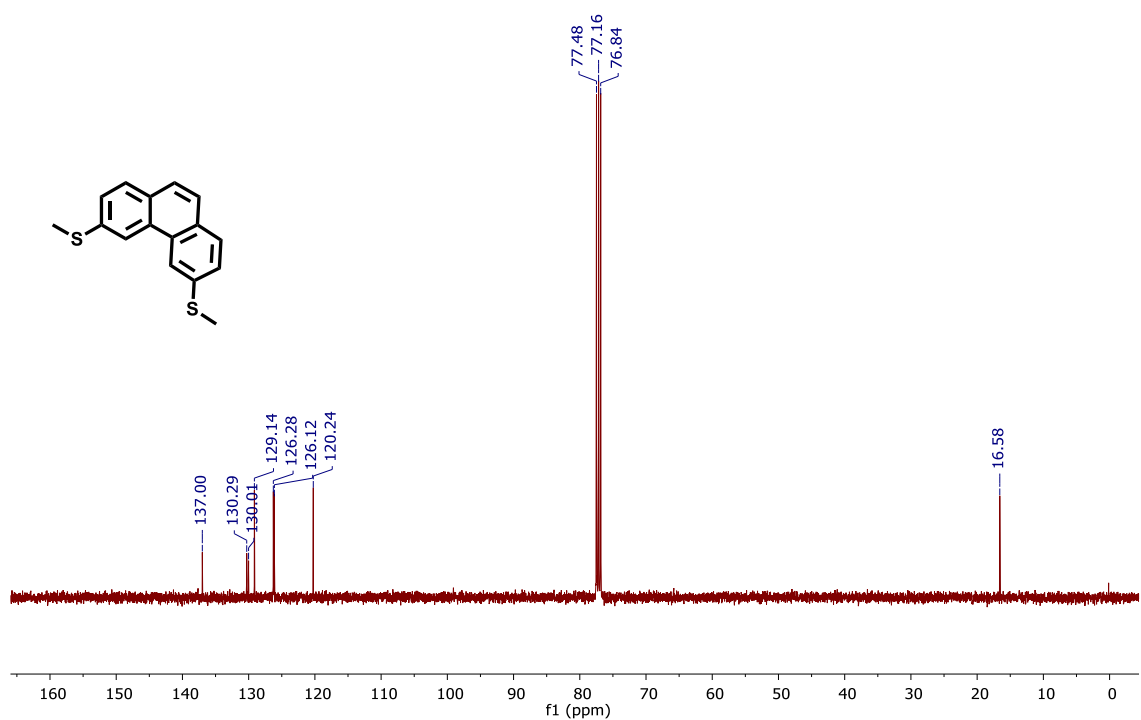

Figure S4. $^1$H NMR (400 MHz) and $^{13}$C NMR (101 MHz) spectra of 3,6-bis(methylthio)phenanthrene (**L-*cis*-OPV2-2SMe**) in CDCl$_3$.

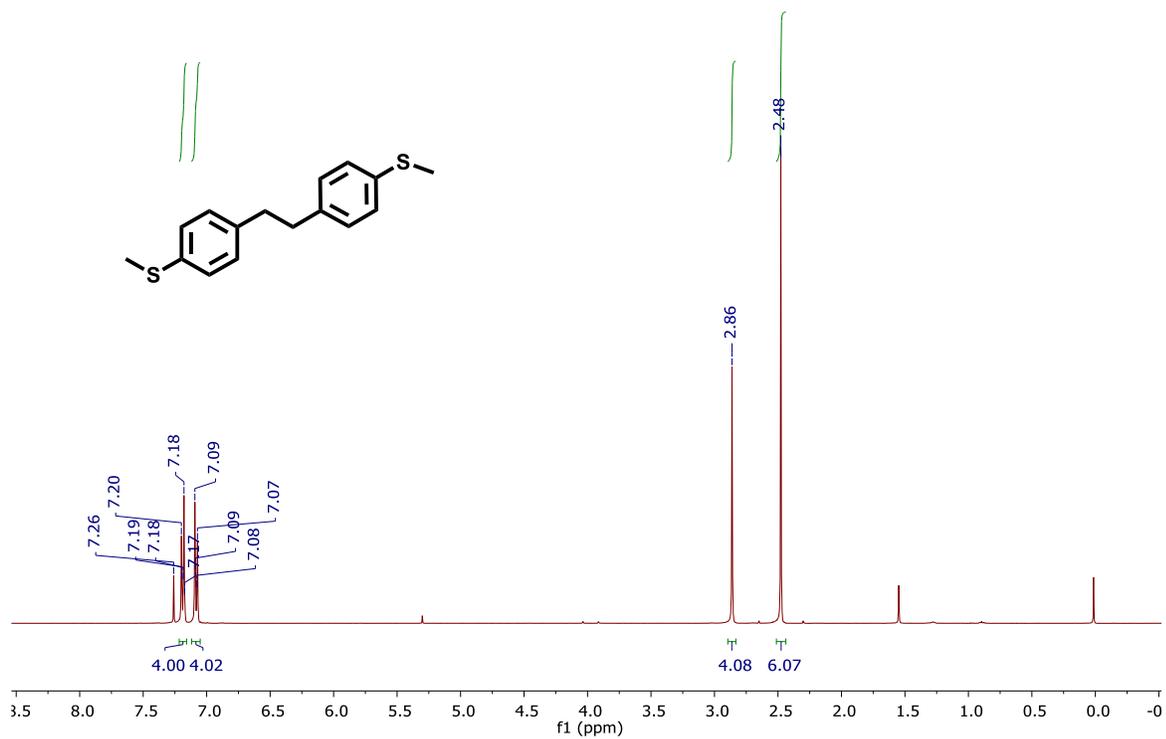
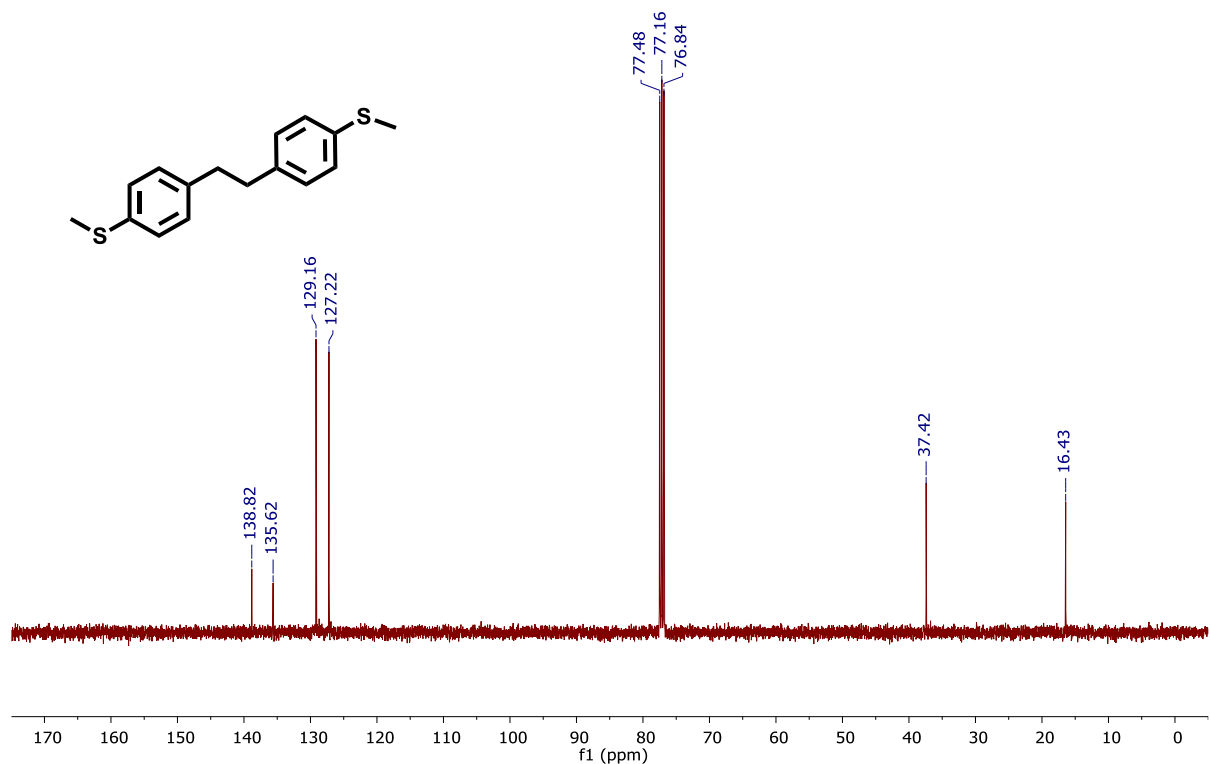

Figure S5. $^1$H NMR (400 MHz) and $^{13}$C NMR (101 MHz) spectra of 1,2-bis(4-(methylthio)phenyl)ethane (**sat-OPV2-2SMe**) in CDCl$_3$.

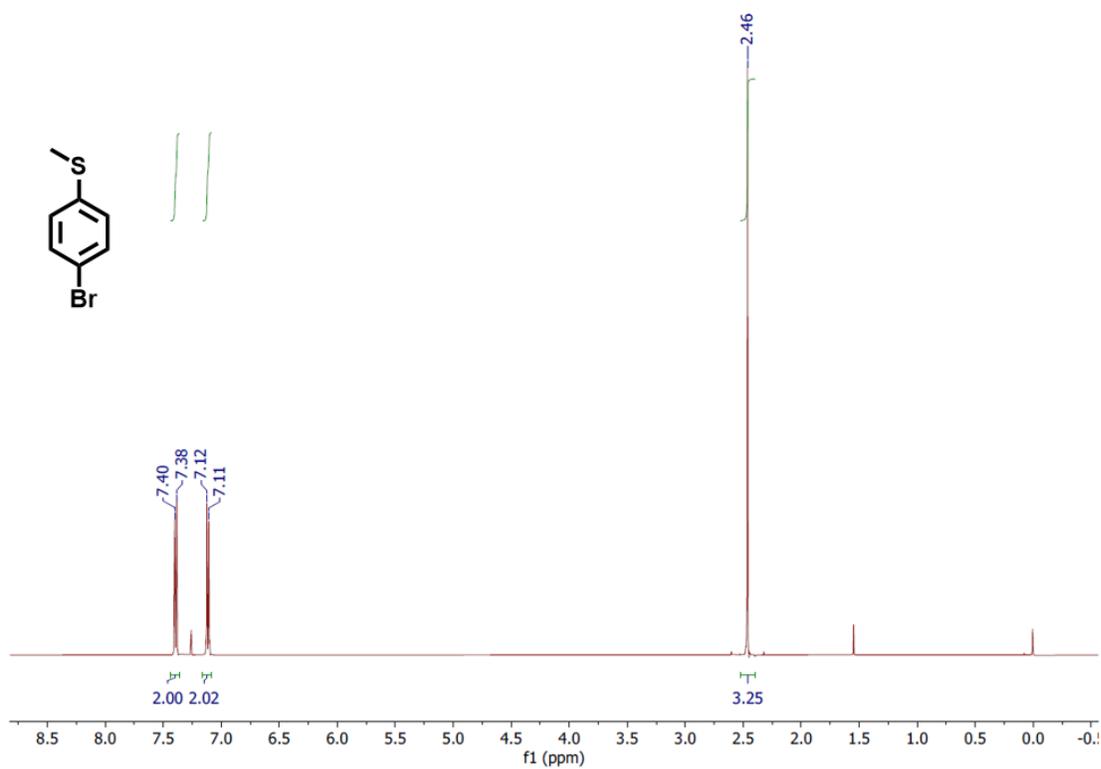

Figure S6. ¹H NMR (500 MHz) spectrum of (4-bromophenyl)(methyl)sulfane (**6**) in CDCl₃.

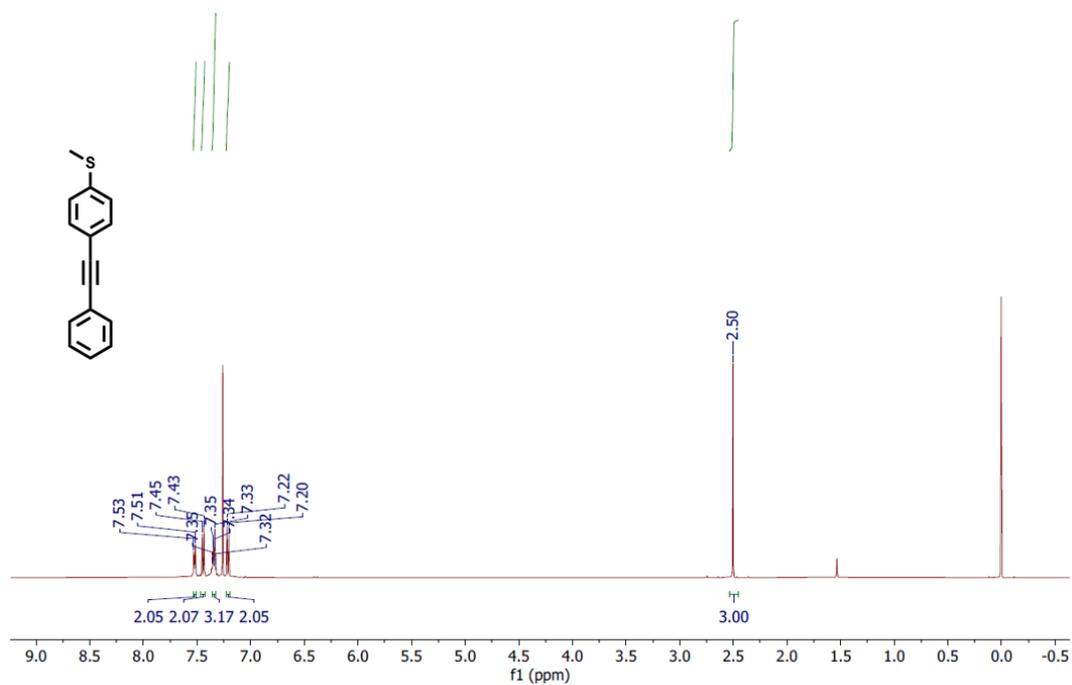

Figure S7. ¹H NMR (500 MHz) spectrum of methyl(4-(phenylethynyl)phenyl)sulfane (**7**) in CDCl₃.

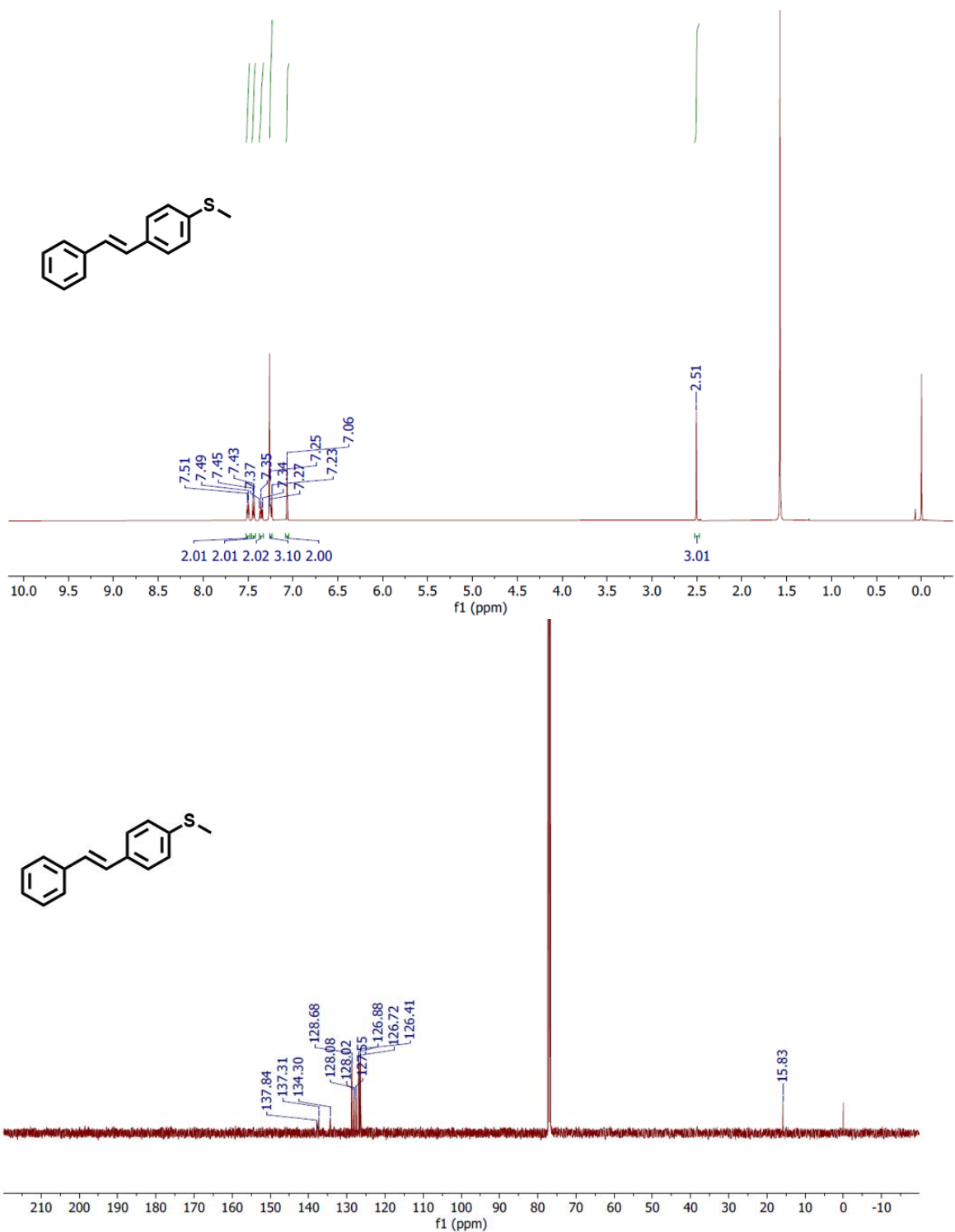

Figure S8. $^1$H NMR (500 MHz) and $^{13}$C NMR (126 MHz) spectra of (*E*)-methyl(4-styrylphenyl)sulfane (***trans*-OPV2-1SMe**) in CDCl$_3$.

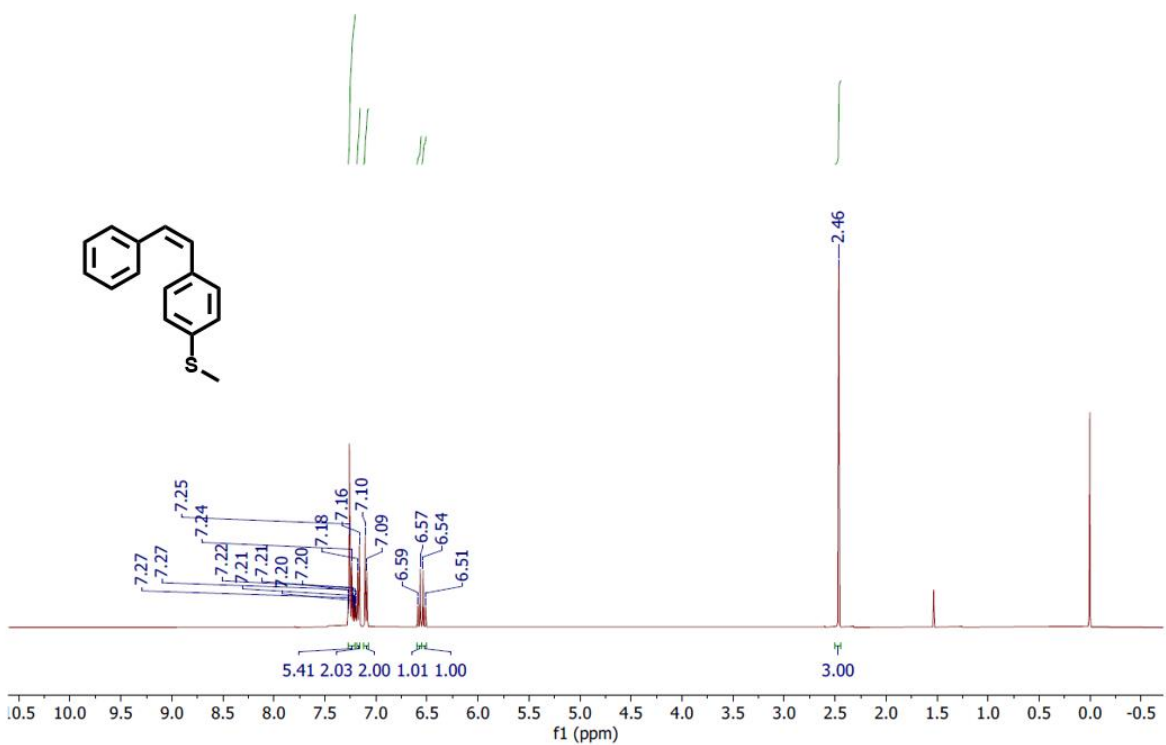
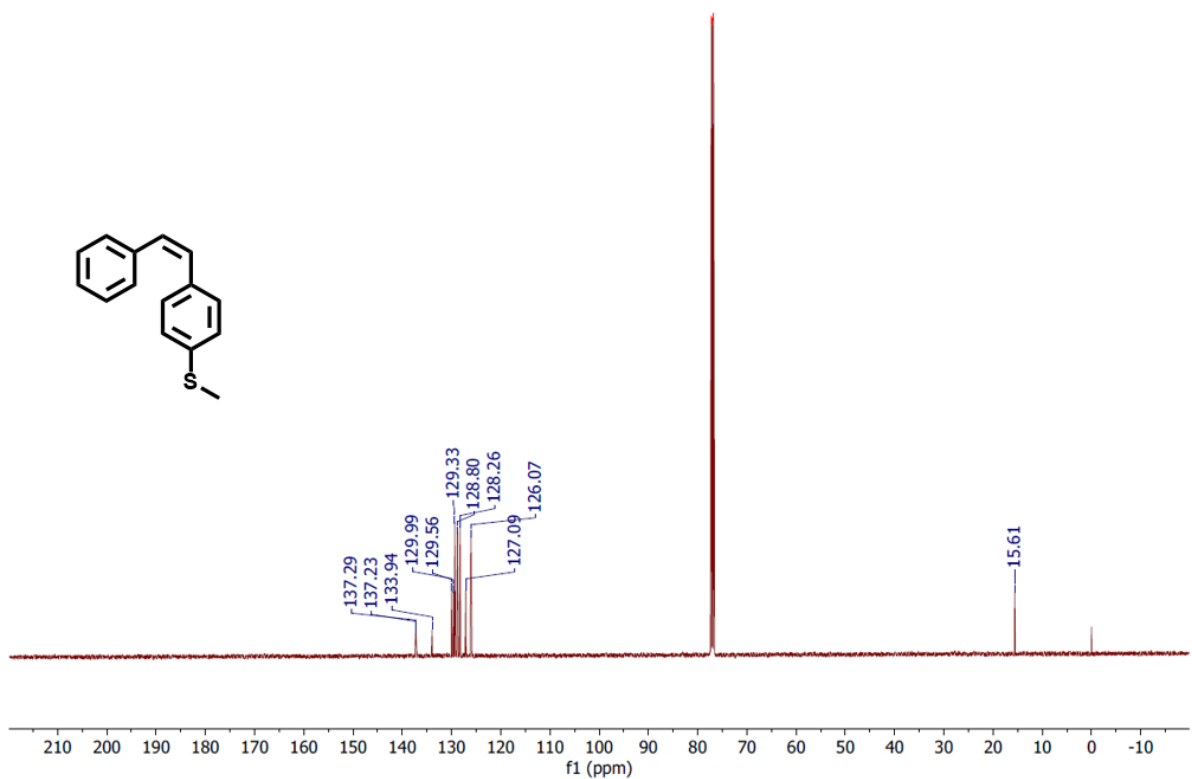

Figure S9. $^1$H NMR (500 MHz) and $^{13}$C NMR (126 MHz) spectra of (Z)-methyl(4-styrylphenyl)sulfane (**cis-OPV2-1SMe**) in CDCl$_3$.

### S.1.3 Mass Spectrometry Characterization

Low resolution mass spectra were recorded on an amaZon SL instrument in the Analytical & Biological Mass Spectrometry Facility, University of Arizona. LRMS (*m/z*): was recorded in the positive/negative mode using $CH_2Cl_2$ in MeOH.

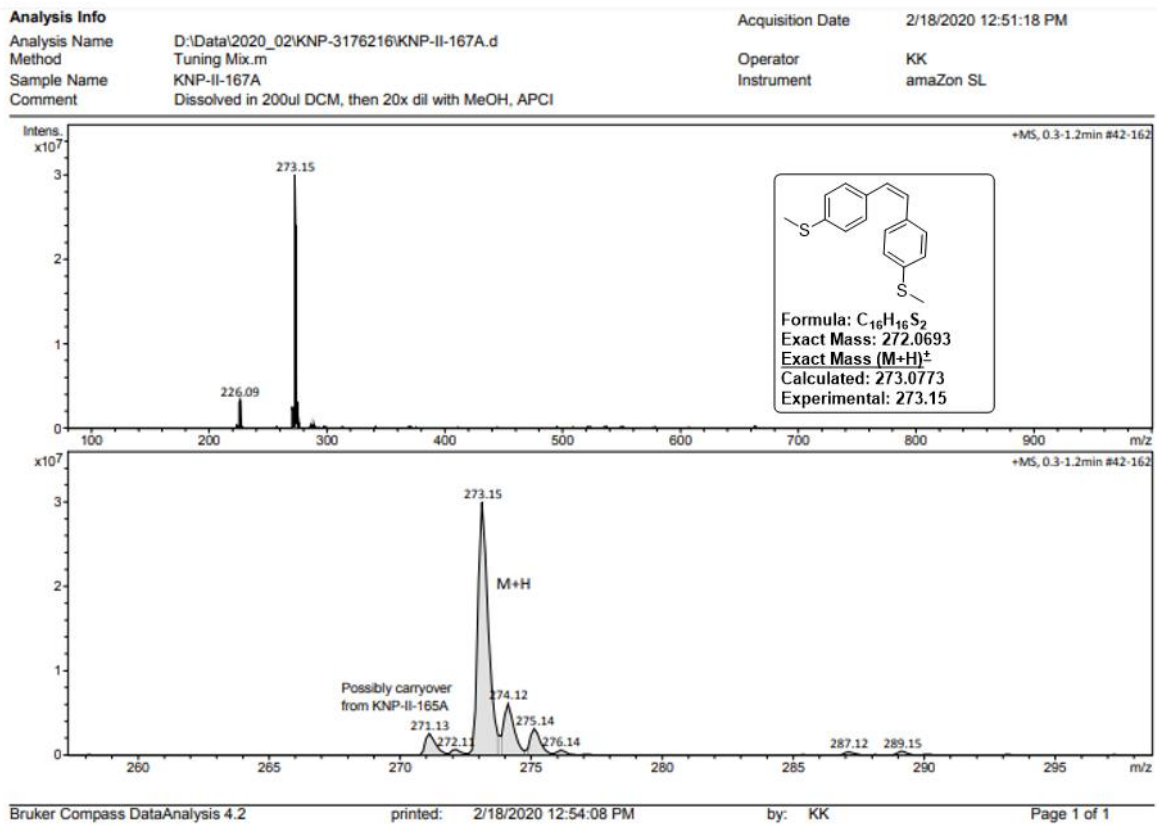

Figure S10. LRMS spectrum of (*Z*)-1,2-bis(4-(methylthio)phenyl)ethene (***cis*-OPV2-2SMe**).

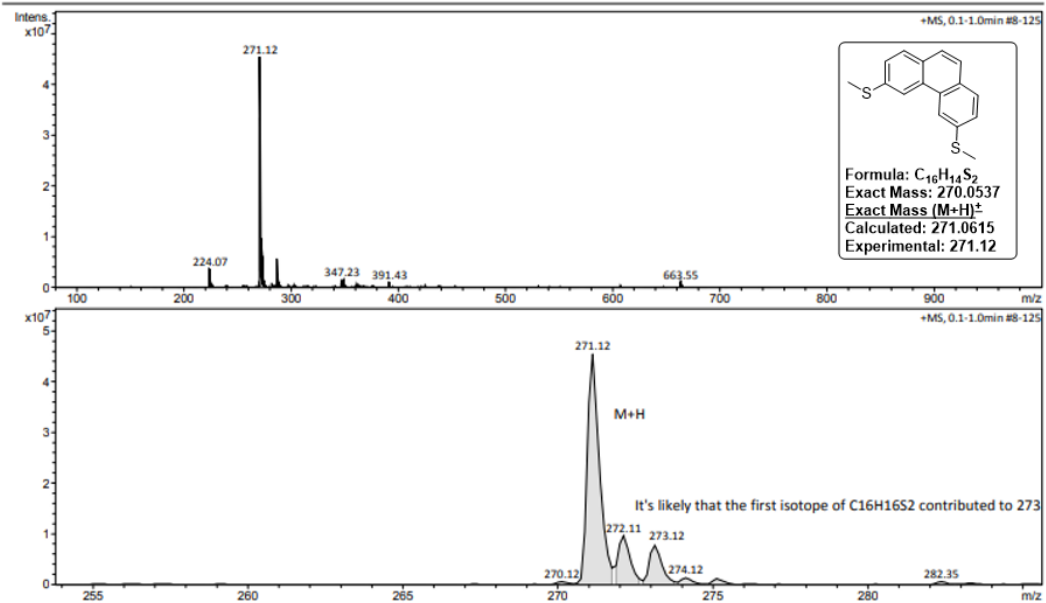

Figure S11. LRMS spectrum of 3,6-bis(methylthio)phenanthrene (**L-*cis*-OPV2-2SMe**).

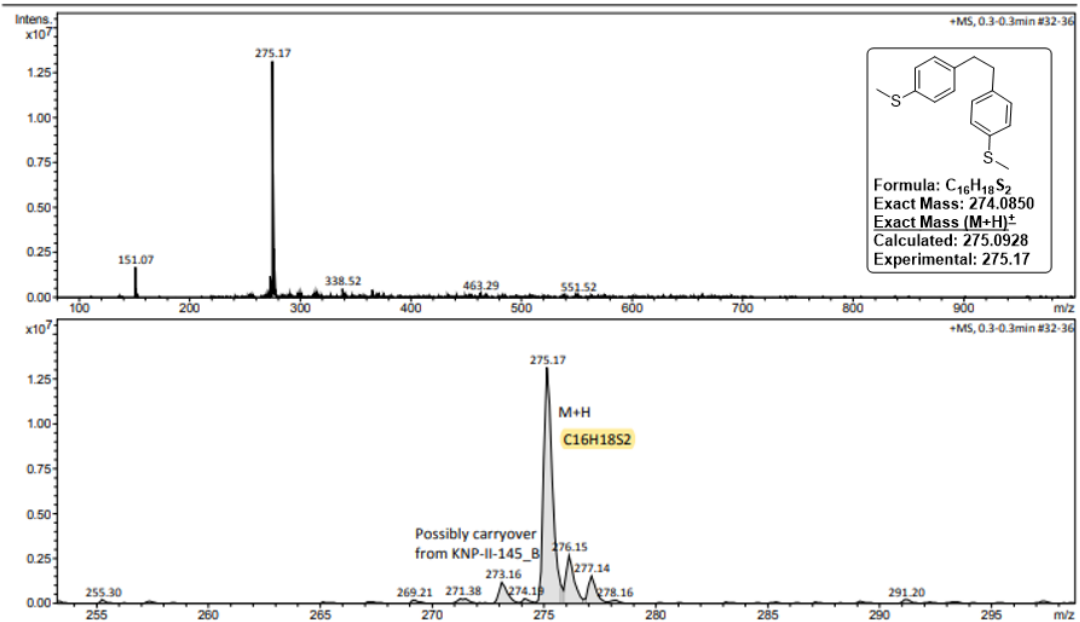

Figure S12. LRMS spectrum of 1,2-bis(4-(methylthio)phenyl)ethane (**sat-OPV2-2SMe**).

## S.2 Datasets Used

The breaking trace data in this work were collected using 15 separate MCBJ samples. For most of these samples, the molecular solution was deposited multiple times ("new deposition") and/or the sample was completely relaxed and then the breaking program was restarted ("starting a new trial"). As in previous work,[17,18] we treat each unique deposition/trial combination as a separate dataset. Datasets with fewer than 1000 traces, as well as datasets that showed no evidence of a molecular feature in the raw 1D conductance histogram, were excluded from analysis.

Details on each of the trans-OPV2-2SMe, cis-OPV2-2SMe, and sat-OPV2-2SMe datasets considered in this work are provided in Table S1, including a unique dataset ID number, an ID number identifying the MCBJ sample used, the molecule name, the number of consecutive traces in the dataset, the concentration of the molecular solution most recently deposited, and the trial and deposition numbers. Each of these datasets was independently clustered for our peak conductance analysis.

The other three molecules (trans-OPV2-1SMe, cis-OPV2-1SMe, and L-cis-OPV2-2SMe) tended to show relatively weak molecular features. Combined with their short lengths, this resulted in relatively small main plateau clusters that were difficult to identify with segment clustering. We therefore chose to group the datasets for these molecules by trial number before clustering them, as shown in Table S2, Table S3, and Table S4. By increasing the absolute number of molecular plateaus in the set of traces clustered, this grouping improved the robustness and ease of identification of main plateau clusters by segment clustering. However, we emphasize that clustering each of the datasets for these three molecules separately would not qualitatively change the main findings of this work; it would simply require more full-valley clusters[17] to be examined to locate the main plateau cluster in each dataset, and it would result in a few cases with ambiguity about which cluster represents the main molecular feature.

The histograms in Figure 2a in the main text are for the datasets in the following tables with the ID numbers (in order): 186, 312, 255, 198, 250, and 279.

Table S1. Details of trans-OPV2-2SMe, cis-OPV2-2SMe, and sat-OPV2-2SMe datasets analyzed in this work.

| Dataset ID# | Sample ID# | Molecule | # of Traces | Concentration (µM) | Trial # | Deposition # |
|---|---|---|---|---|---|---|
| 186 | 114-5 | trans-OPV2-2SMe | 2594 | 1 | 1 | 3 |
| 189 | 114-5 | trans-OPV2-2SMe | 2543 | 1 | 2 | 8 |
| 190 | 114-5 | trans-OPV2-2SMe | 8605 | 10 | 2 | 1[a] |
| 191 | 114-5 | trans-OPV2-2SMe | 9236 | 10 | 2 | 2 |
| 194 | 125-3 | trans-OPV2-2SMe | 6353 | 10 | 1 | 1[b] |
| 195 | 125-3 | trans-OPV2-2SMe | 2837 | 10 | 1 | 2 |
| 198 | 118-3 | cis-OPV2-2SMe | 9257 | 10 | 1 | 1[b] |
| 199 | 118-3 | cis-OPV2-2SMe | 6822 | 10 | 1 | 2 |
| 200 | 118-3 | cis-OPV2-2SMe | 2257 | 10 | 1 | 3 |
| 201 | 118-3 | cis-OPV2-2SMe | 4543 | 10 | 2 | 3 |
| 202 | 118-3 | cis-OPV2-2SMe | 6906 | 10 | 2 | 4 |
| 207 | 125-5 | cis-OPV2-2SMe | 3560 | 1 | 2 | 3 |
| 208 | 125-5 | cis-OPV2-2SMe | 4627 | 1 | 2 | 5 |

| | | | | | | | |
|---|---|---|---|---|---|---|---|
| 209 | 125-5 | cis-OPV2-2SMe | 4499 | 10 | 2 | 1[c] |
| 210 | 125-5 | cis-OPV2-2SMe | 2955 | 10 | 3 | 1[c] |
| 253 | 127-4 | sat-OPV2-2SMe | 5690 | 1 | 2 | 1 |
| 254 | 127-4 | sat-OPV2-2SMe | 2376 | 1 | 2 | 2 |
| 255 | 127-4 | sat-OPV2-2SMe | 5741 | 1 | 2 | 3 |
| 256 | 127-4 | sat-OPV2-2SMe | 2046 | 1 | 3 | 3 |
| 257 | 127-4 | sat-OPV2-2SMe | 2003 | 1 | 3 | 4 |
| 259 | 133-3 | sat-OPV2-2SMe | 2702 | 1 | 4 | 2 |
| 260 | 133-3 | sat-OPV2-2SMe | 4163 | 1 | 4 | 3 |

[a] First 10 µM deposition, preceded by eight 1 µM depositions.
[b] First 10 µM deposition, preceded by two 1 µM depositions.
[c] First 10 µM deposition, preceded by five 1 µM depositions.

Table S2. Details of trans-OPV2-1SMe datasets analyzed in this work. Datasets were grouped according to MCBJ sample and trial number before clustering, as indicated by the first column.

| Trial Combination ID# | Dataset ID# | Sample ID# | Molecule | # of Traces | Concentration (µM) | Trial # | Deposition # |
|---|---|---|---|---|---|---|---|
| C1 | 265 | 135-4 | trans-OPV2-1SMe | 1663 | 10 | 3 | 1[a] |
| C1 | 266 | 135-4 | trans-OPV2-1SMe | 3022 | 10 | 3 | 3 |
| C1 | 267 | 135-4 | trans-OPV2-1SMe | 1741 | 10 | 3 | 4 |
| C2 | 268 | 135-4 | trans-OPV2-1SMe | 2189 | 10 | 4 | 4 |
| C2 | 269 | 135-4 | trans-OPV2-1SMe | 2905 | 10 | 4 | 5 |
| N/A | 273 | 127-2 | trans-OPV2-1SMe | 1560 | 10 | 4 | 1[b,c] |
| C3 | 309 | 131-3 | trans-OPV2-1SMe | 4716 | 10 | 2 | 1 |
| C3 | 310 | 131-3 | trans-OPV2-1SMe | 1979 | 10 | 2 | 2 |
| C4 | 311 | 131-3 | trans-OPV2-1SMe | 2230 | 10 | 3 | 2 |
| C4 | 312 | 131-3 | trans-OPV2-1SMe | 5258 | 10 | 3 | 3 |
| C5 | 313 | 131-3 | trans-OPV2-1SMe | 284 | 10 | 4 | 3 |
| C5 | 314 | 131-3 | trans-OPV2-1SMe | 4568 | 10 | 4 | 4 |

[a] First 10 µM deposition, preceded by one 1 µM deposition.
[b] First 10 µM deposition, preceded by two 1 µM depositions.
[c] This dataset is a subset of an entire trial/deposition block, in order to exclude an unreproducible noise feature that occurred partway through.

Table S3. Details of cis-OPV2-1SMe datasets analyzed in this work. Datasets were grouped according to MCBJ sample and trial number before clustering, as indicated by the first column.

| Trial Combination ID# | Dataset ID# | Sample ID# | Molecule | # of Traces | Concentration (µM) | Trial # | Deposition # |
|---|---|---|---|---|---|---|---|
| N/A | 325 | 126-1 | cis-OPV2-1SMe | 3263 | 10 | 2 | 1[a] |
| C6 | 326 | 126-1 | cis-OPV2-1SMe | 2542 | 10 | 3 | 2 |
| C6 | 327 | 126-1 | cis-OPV2-1SMe | 2922 | 10 | 3 | 3 |
| C6 | 328 | 126-1 | cis-OPV2-1SMe | 1491 | 10 | 3 | 4 |
| C7 | 329 | 126-1 | cis-OPV2-1SMe | 2946 | 10 | 4 | 4 |
| C7 | 330 | 126-1 | cis-OPV2-1SMe | 5576 | 10 | 4 | 5 |
| C8 | 240 | 133-5 | cis-OPV2-1SMe | 2229 | 1 | 1 | 1 |
| C8 | 241 | 133-5 | cis-OPV2-1SMe | 3528 | 1 | 1 | 2 |

| | 242 | 133-5 | cis-OPV2-1SMe | 5069 | 10 | 1 | 1[b] |
| --- | --- | --- | --- | --- | --- | --- | --- |
| | 243 | 133-5 | cis-OPV2-1SMe | 3374 | 10 | 1 | 2 |
| N/A | 244 | 133-5 | cis-OPV2-1SMe | 3559 | 10 | 2 | 3 |
| | 247 | 135-2 | cis-OPV2-1SMe | 4508 | 10 | 1 | 3 |
| C9 | 248 | 135-2 | cis-OPV2-1SMe | 2252 | 10 | 1 | 4 |
| | 249 | 135-2 | cis-OPV2-1SMe | 3650 | 10 | 1 | 4[c] |
| N/A | 250 | 135-2 | cis-OPV2-1SMe | 6766 | 10 | 2 | 5 |

[a] First 10 µM deposition, preceded by two 100 nM depositions.
[b] First 10 µM deposition, preceded by two 1 µM depositions.
[c] Pure DCM was deposited between the previous dataset and this one.

Table S4. Details of L-cis-OPV2-2SMe datasets analyzed in this work. Datasets were grouped according to MCBJ sample and trial number before clustering, as indicated by the first column.

| Trial Combination ID# | Dataset ID# | Sample ID# | Molecule | # of Traces | Concentration (µM) | Trial # | Deposition # |
| --- | --- | --- | --- | --- | --- | --- | --- |
| | 277 | 128-2 | L-cis-OPV2-2SMe | 5897 | 10 | 3 | 2 |
| C10 | 278 | 128-2 | L-cis-OPV2-2SMe | 1502 | 10 | 3 | 3 |
| | 279 | 128-2 | L-cis-OPV2-2SMe | 3281 | 10 | 3 | 4 |
| C11 | 280 | 128-2 | L-cis-OPV2-2SMe | 2662 | 10 | 4 | 4 |
| | 281 | 128-2 | L-cis-OPV2-2SMe | 2647 | 10 | 4 | 5 |
| C12 | 284 | 133-4 | L-cis-OPV2-2SMe | 1848 | 10 | 1 | 1 |
| | 285 | 133-4 | L-cis-OPV2-2SMe | 2503 | 10 | 1 | 2 |
| | 286 | 133-4 | L-cis-OPV2-2SMe | 3986 | 10 | 3 | 2 |
| C13 | 287 | 133-4 | L-cis-OPV2-2SMe | 4123 | 10 | 3 | 3 |
| | 288 | 133-4 | L-cis-OPV2-2SMe | 1730 | 10 | 3 | 4 |
| | 289 | 133-4 | L-cis-OPV2-2SMe | 2773 | 10 | 3 | 5 |
| C14 | 290[a] | 133-4 | L-cis-OPV2-2SMe | 1136 | 10 | 4 | 5 |
| | 291[a] | 133-4 | L-cis-OPV2-2SMe | 3155 | 10 | 4 | 6 |
| C15 | 296 | 120-5 | L-cis-OPV2-2SMe | 5616 | 10 | 1 | 1 |
| | 297 | 120-5 | L-cis-OPV2-2SMe | 3454 | 10 | 1 | 2 |
| C16 | 298 | 120-5 | L-cis-OPV2-2SMe | 4012 | 10 | 2 | 2 |
| | 299 | 120-5 | L-cis-OPV2-2SMe | 4338 | 10 | 2 | 3 |
| C17 | 300 | 120-5 | L-cis-OPV2-2SMe | 1220 | 10 | 3 | 3 |
| | 301 | 120-5 | L-cis-OPV2-2SMe | 4236 | 10 | 3 | 4 |

[a] The piezo pushing speed was set to 30 µm/s for these two datasets instead of our standard choice of 60 µm/s.

### S.3 Comment on cis-OPV2-1SMe Datasets

Whereas for the other five molecules, the raw 1D conductance histograms for each dataset look quite similar to the examples shown in Figure 2 in the main text, for cis-OPV2-1SMe some datasets display a broader and flatter feature without a clear peak (see additional examples in Figure S13). It is thus possible that we are experimentally measuring a broader mixture of binding conformations for cis-OPV2-1SMe than for the other OPV2 molecules, and this possibility is an important caveat to keep in mind in this work. However, we also note that the main plateau clusters identified by segment clustering in these flatter cis-OPV2-1SMe distributions agree well with both the clusters identified in the peaked cis-OPV2-1SMe distributions and the location of those raw data peaks themselves (SI section S.4). Therefore, we suggest that the feature identified by segment clustering in the cis-OPV2-1SMe datasets represents molecular transport in this molecule, and is suitable for comparison to the features identified for the other OPV2 constructs. The main text of this manuscript proceeds under this assumption.

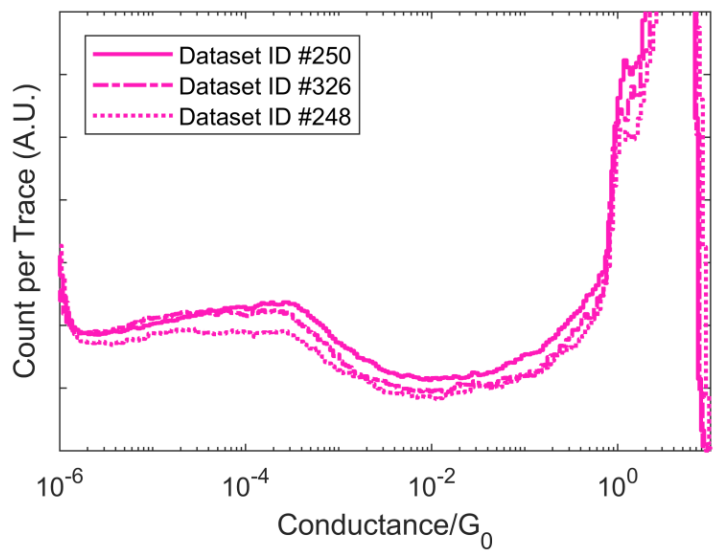

Figure S13. Overlaid raw 1D histograms for three different cis-OPV2-1SMe datasets. Whereas the example dataset shown in Figure 2a in the main text (ID #250, included here as well) shows a subtle peak at ~$10^{-3.6}$ $G_0$, some of the other datasets for this molecule instead have a much broader and flatter feature with no obvious peak. Dataset ID numbers refer to Table S3.

## S.4 Use of Segment Clustering

Segment clustering was applied to each dataset (or dataset combination) from Table S1, Table S2, Table S3, and Table S4 in the manner described in Bamberger et al.[17] In each case, a single "main plateau cluster" corresponding to relatively flat trace segments concentrated in the vicinity of the conductance peak in the raw 1D conductance histogram was unambiguously identified. Due to the short length of these molecular features (as well as some of them being quite weak), the minimum valley size described in Bamberger et al. needed to be lowered to 0.25% (or slightly lower in a few cases) in order to identify these clusters.

Continuing to follow the method of Bamberger et al., each dataset was clustered 12 separate times with different values of the *minPts* parameter. The main plateau cluster was identified in one of these outputs, and then the analogous cluster was identified from the other 11 outputs using an automated algorithm. Histograms were created using all conductance data points from all trace segments assigned to each main plateau cluster, and then unrestricted Gaussian fits were performed to determine a single peak conductance value for each clustering output. Figure S14a summarizes these peak conductance values for each clustered dataset, with the points representing the median from among the 12 different values for each dataset and the error bars representing the range of the middle 8 out of 12 values.

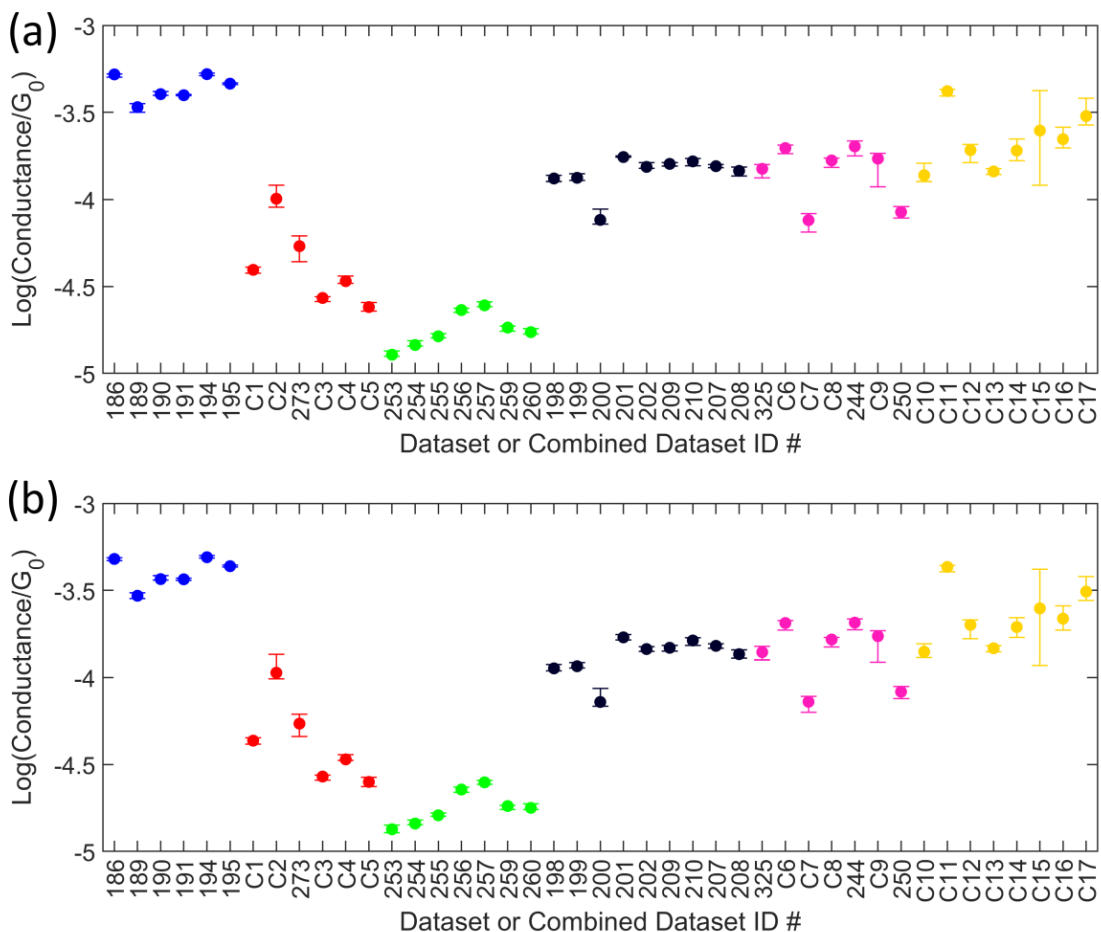

Figure S14. (a) Summary of the peak conductance values for the identified main plateau clusters of each dataset, determined using unrestricted Gaussian fits. Each point represents the median from among the 12 clustering outputs produced for each dataset and the error bars represent the range of the middle 8 of those 12 values. The ID numbers correspond to Table S1, Table S2, Table S3, and Table S4, and the points are color-coded by molecule following the scheme from the main text. (b) The same as (a), but using median conductance values rather than Gaussian fits to determine the "peak" conductance value for each main plateau cluster.

While a Gaussian shape fit the distribution of conductances for each main plateau cluster reasonably well, many of these distributions displayed meaningful deviations from the fit, sometimes including small satellite peaks. Therefore, to account for the possibility that Gaussian fits may not be the best choice for summarizing the conductance distribution for each main plateau cluster, we also looked at the median of each distribution. As shown in Figure S14b, the results of using conductance medians are almost identical to those from using Gaussian fits, confirming that we are calculating reasonable summaries of these data.

## S.5 Apparent Molecular Lengths

To compare the apparent lengths of the different molecular features, for each identified main plateau cluster we calculated the median value of the end points of all segments in that cluster.

Box plots summarizing the set of such end point medians for all 12 clustering outputs for all datasets collected for a given molecule are shown in Figure S15.

While there is a fair amount of variation between datasets that makes precise length-measurements difficult, the distributions in Figure S15 clearly show systematically longer lengths for the trans and saturated molecules than for the cis molecules, as well as no obvious differences within each of those groups. In particular, the junction lengths of the single-linker cis and trans molecules are very similar to those of their double-linker analogues. This suggests that the single-linker molecules are forming direct Au-π linkages on one side rather than creating stacked two-molecule bridges, which would be expected to appear significantly longer than the double-linker junctions.[19]

When comparing the distributions in Figure S15 to the expected lengths of each molecule, it is important to note that: 1) a snap-back distance of ~0.5 nm should be added to the apparent lengths;[20,21] and 2) some molecules likely break off before full extension, so the *median* end point can be expected to underestimate the fully-extended molecular junction length.

One potential reason for the significant variation in the median end points for each molecule is that there may be systematic errors in the attenuation ratio used to convert piezo movements into inter-electrode distance. This ratio is calculated once for each MCBJ sample before molecular deposition and then assumed to remain constant,[17] but it may vary over time due to, e.g., changes in the elasticity of the substrate with repeated use.

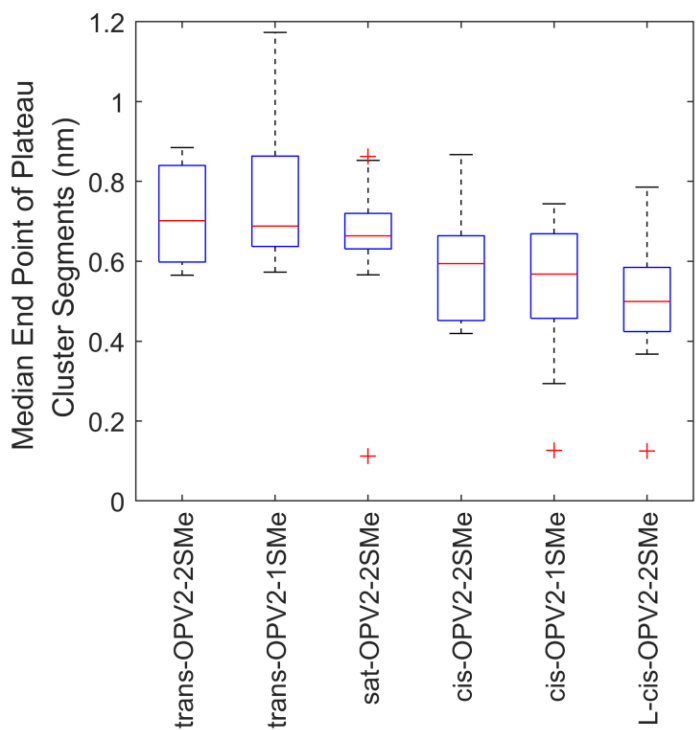

Figure S15. Box plots summarizing the set of median segment end points from the set of all main plateau clusters identified for each molecule. The red line represents the median of each distribution, the blue box represents the inter-quartile range, and the whiskers extend to the farthest points not considered outliers. Outliers (red crosses) are defined as points farther than 1.5 times the interquartile range from the bottom or top of each box.

## S.6 Preparation of Two-Probe Geometries for NEGF-DFT Calculations

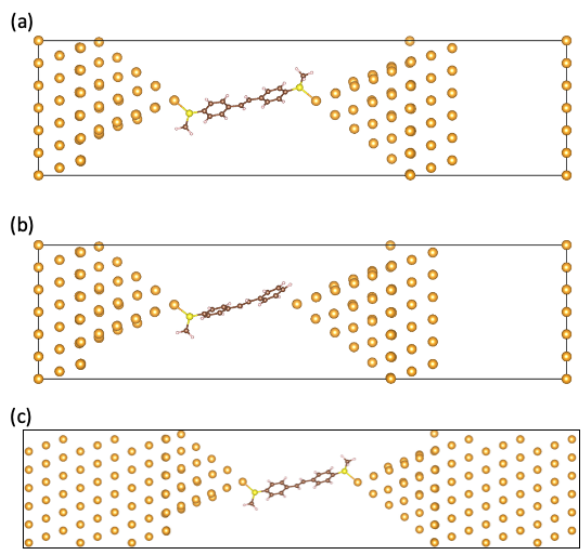

Figure S16. The structure relaxations of the in-junction geometries were carried out on systems as shown in (a) for the 2SMe systems and (b) for the 1SMe systems. Once the minimum energy Au-Au gap separation was determined, the Au electrodes were extended to form the two-probe geometry shown in (c).

## S.7 NEGF-DFT Results for Single-Linker Molecules

In order to assess the binding motifs of cis- and trans-OPV2-1SMe in the junction, we carried out electronic structure optimizations and transport calculations as a function of gap size. Figure S17 shows the result of these calculations for trans-OPV2-1SMe together with the total energy relative to relaxed minimum at 44.8 Å. As is evident, there are a number of energetically close-lying structural minima with somewhat different transmission functions. This makes establishing a close correspondence to likely experimentally observed in-junction structures challenging. Moreover, the overriding importance of long-range van der Waals interactions is not accurately captured at the level of theory used here.

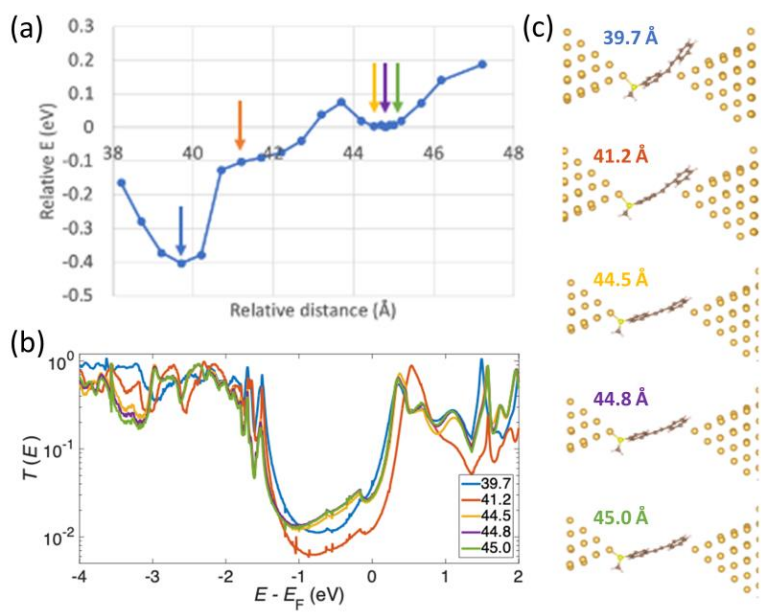

Figure S17. Computational results for trans-OPV2-1SMe. (a) Relative energies of optimized in-junction molecular structures for varying gap sizes measured relative to an absolute minimum at 44.8 Å. (b) Transmission functions for five of the structures in (a), color-coded to the points indicated in (a) by the arrows. (c) Structural views of the five structures indicated with colored arrows in (a).

### S.8 Barrier to -SMe Rotation

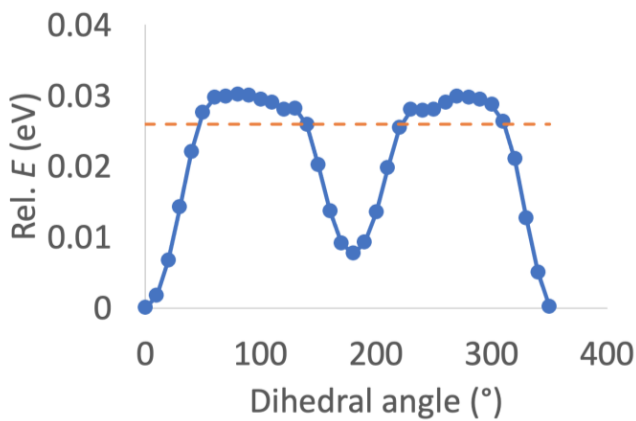

Figure S18. Potential energy surface for the rotation of one -SMe linker with respect to its adjacent phenyl ring of the isolated trans-OPV2-2SMe molecule (with the two phenyl rings kept in the same plane, but all atoms free to move otherwise), calculated with Orca[22] using DFT at the B3LYP[23,24]/6-31G(d) level. The barriers to rotation are comparable to $k_B T$ at room temperature (red dashed line), suggesting that these linkers easily sample many different conformations during transport experiments.

## S.9 Gas-Phase Ionization Energies

Because transport is thought to be dominated by the HOMO channel for the molecules considered in this work, single-molecule conductance is expected to depend in part on the molecular ionization energies, with larger values implying larger hole-injection barriers and hence lower conductance.[25–27] To gain qualitative insight into this dependence, we thus calculated gas-phase vertical ionization energies for each of the six molecules considered in this work. Calculations were performed using Gaussian 16[28] using the B3LYP functional and the 6-311++G(p,d) basis set. For most of these molecules, multiple local minima structures were discovered depending on the orientation of the -SMe linker groups. In order to be comparable across the series, we focus here on the structures in which both -SMe linkers are in plane with their neighboring phenyl rings and the methyl groups are pointed away from the center of the molecule.

Table S5. Gas-phase vertical ionization energies calculated for the six molecules considered in this work with their -SMe linkers in-plane and pointed outwards.

| Molecule | Gas-Phase Vertical Ionization Energy (eV) |
|---|---|
| trans-OPV2-2SMe | 6.59 |
| trans-OPV2-1SMe | 6.91 |
| sat-OPV2-2SMe | 7.00 |
| cis-OPV2-2SMe | 6.76 |
| cis-OPV2-1SMe | 7.09 |
| L-cis-OPV2-2SMe | 6.79 |

## S.10 Controlling Au-S Binding Angles Through Gap Size

In Figure 6a in the main text, both Au-S-C-C dihedral angles are varied for L-cis-OPV2-2SMe by introducing artificial constraints. While that approach has the advantage of directly controlling the parameter of interest, it also has the limitation that the molecule may be forced into quite strained and unrealistic overall conformations. Here we thus present a complementary approach in which the size of the nanogap is varied, which is accompanied by steady changes in the two Au-S-C-C dihedral angles as the molecule is allowed to relax at each fixed gap size (see legend of Figure S19). This latter approach has the advantage of allowing the molecule to adopt more realistic conformations, but at the price of introducing a second variable (gap size) that is changing at the same time.

As shown in Figure S19, this second approach produces the same qualitative trends as observed in Figure 6a in the main text, thus suggesting clearly that those trends are indeed a result of the Au-S-C-C angle changes rather than an artefact of how the calculations were performed.

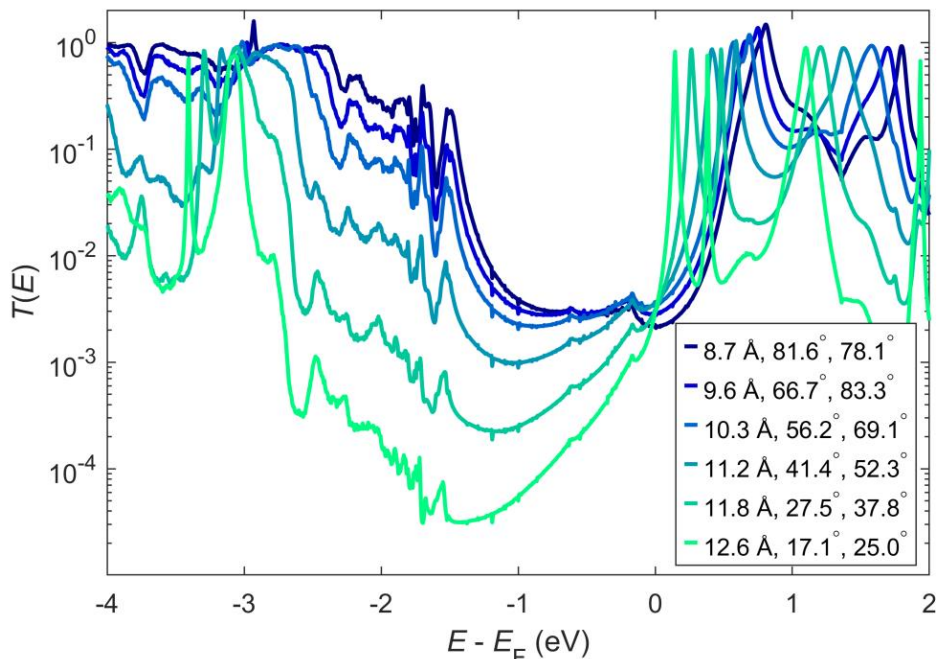

Figure S19. Overlaid transmission functions calculated for relaxed geometries of L-cis-OPV2-2SMe at different Au-Au gap sizes. The gap size and the two Au-S-C-C dihedral angles are listed in the legend, demonstrating that increasing the gap size is correlated with rotating these angles from 90° towards 0°. The qualitative changes to the transmission function thus mirror those seen in Figure 6a in the main text, in which the Au-S-C-C dihedral angles were controlled directly.

## S.11   Electronic Structure Changes During -SMe Rotation

In Figure 6 in the main text, the HOMO and LUMO peaks in the L-cis-OPV2-2SMe transmission function are observed to both shift towards more positive energy as the Au-S-C-C angles are varied from 0° towards 90°. To demonstrate that this effect is primarily a result of changes to the molecule itself rather than molecule/metal coupling, single-point gas phase DFT calculations were performed after removing all gold atoms for each of the molecular structures in Figure 6. As shown in Figure S20, the calculated gas-phase ionization energies and electron affinities for L-cis-OPV2-2SMe both decrease by 100s of meV as the Au-S-C-C angles are rotated from 0° towards 90°, consistent with the transmission peak shifts in Figure 6 of the main text.

The likely reason for this shift of the transport levels is that in the junction the Au-S-C-C dihedral angles are strongly anticorrelated with the S-C-C-C dihedral angle defining the rotation of the -SMe linker relative to its attached phenyl ring (Figure S21). This means that when the Au-S bond is mostly perpendicular to the phenyl ring, the -SMe linker is mostly planar with it, and vice versa. Whereas the -SMe linkers have little effect on molecular electronic structure in their perpendicular orientation, when they are planar with the rest of the molecular π-system they act as electron donating substituents, raising the molecular energy levels relative to vacuum. This same effect of substituent orientation on electron donating character has been previously observed for an -OMe substituent,[25] but the fact that linker groups can also act as electron donation/withdrawing substituents is rarely considered.

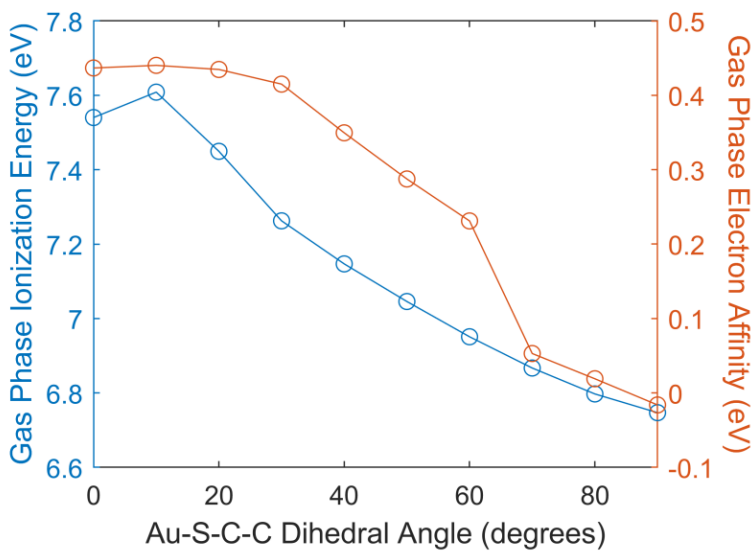

Figure S20. Gas-phase ionization energies (blue) and electron affinities (orange) determined from single-point DFT calculations for each of the ten structures of L-cis-OPV2-2SMe from Figure 6 in the main text. This demonstrates that even without the presence of the gold electrodes, the molecular transport levels shift to more negative energies as the Au-S-C-C dihedral angles are rotated from 90° towards 0° (*larger* ionization energy and electron affinity values imply the transport levels are *lower* in energy relative to vacuum).

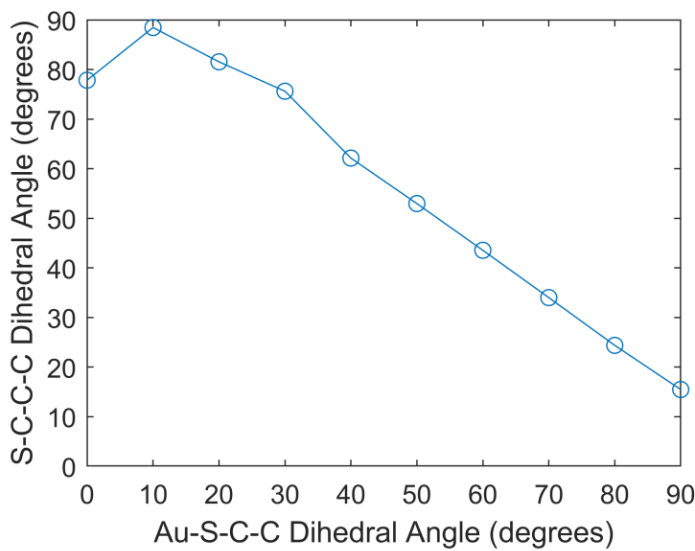

Figure S21. Relationship between the Au-S-C-C dihedral angle (defined in Figure 3 in the main text) and the S-C-C-C dihedral angle that defines the rotation of the -SMe methyl group relative to the attached phenyl ring (average of left and right sides of molecule) for the ten structures in Figure 6 in the main text. This demonstrates that changing the geometry of the gold-molecule linkage simultaneously causes changes to the internal electronic structure of the molecule.

# References

(1) Chu, C.-J.; Wu, G.-S.; Ma, H.-I.; Venkatesan, P.; Thirumalaivasan, N.; Wu, S.-P. A Fluorescent Turn-on Probe for Detection of Hypochlorus Acid and Its Bioimaging in Living Cells. *Spectrochimica Acta Part A: Molecular and Biomolecular Spectroscopy* **2020**, *233*, 118234. https://doi.org/10.1016/j.saa.2020.118234.

(2) Delcaillau, T.; Boehm, P.; Morandi, B. Nickel-Catalyzed Reversible Functional Group Metathesis between Aryl Nitriles and Aryl Thioethers. *Journal of the American Chemical Society* **2021**, *143* (10), 3723–3728. https://doi.org/10.1021/jacs.1c00529.

(3) Lim, B.; Jung, H.; Yoo, H.; Park, M.; Yang, H.; Chung, W.-J.; Koo, S. Synthetic Strategy for Tetraphenyl-Substituted All- E -Carotenoids with Improved Molecular Properties. *European Journal of Organic Chemistry* **2020**, *2020* (11), 1769–1777. https://doi.org/10.1002/ejoc.202000130.

(4) Ivie, J. A.; Bamberger, N. D.; Parida, K. N.; Shepard, S.; Dyer, D.; Saraiva-Souza, A.; Himmelhuber, R.; McGrath, D. V.; Smeu, M.; Monti, O. L. A. Correlated Energy-Level Alignment Effects Determine Substituent-Tuned Single-Molecule Conductance. *ACS Applied Materials & Interfaces* **2021**, *13* (3), 4267–4277. https://doi.org/10.1021/acsami.0c19404.

(5) Moorthy, J. N.; Mandal, S.; Mukhopadhyay, A.; Samanta, S. Helicity as a Steric Force: Stabilization and Helicity-Dependent Reversion of Colored o -Quinonoid Intermediates of Helical Chromenes. *Journal of the American Chemical Society* **2013**, *135* (18), 6872–6884. https://doi.org/10.1021/ja312027c.

(6) Bamberger, N. D.; Dyer, D.; Parida, K. N.; McGrath, D. V.; Monti, O. L. A. Grid-Based Correlation Analysis to Identify Rare Quantum Transport Behaviors. *The Journal of Physical Chemistry C* **2021**, *125* (33), 18297–18307. https://doi.org/10.1021/acs.jpcc.1c04794.

(7) Park, G.; Yi, S. Y.; Jung, J.; Cho, E. J.; You, Y. Mechanism and Applications of the Photoredox Catalytic Coupling of Benzyl Bromides. *Chemistry - A European Journal* **2016**, *22* (49), 17790–17799. https://doi.org/10.1002/chem.201603517.

(8) Arakawa, Y.; Kang, S.; Tsuji, H.; Watanabe, J.; Konishi, G. The Design of Liquid Crystalline Bistolane-Based Materials with Extremely High Birefringence. *RSC Advances* **2016**, *6* (95), 92845–92851. https://doi.org/10.1039/C6RA14093A.

(9) Handa, S.; Smith, J. D.; Zhang, Y.; Takale, B. S.; Gallou, F.; Lipshutz, B. H. Sustainable HandaPhos- Ppm Palladium Technology for Copper-Free Sonogashira Couplings in Water under Mild Conditions. *Organic Letters* **2018**, *20* (3), 542–545. https://doi.org/10.1021/acs.orglett.7b03621.

(10) Miyaura, N.; Suzuki, A. PALLADIUM-CATALYZED REACTION OF 1-ALKENYLBORONATES WITH VINYLIC HALIDES: (1Z,3E)-1-PHENYL-1,3-OCTADIENE. *Organic Syntheses* **1990**, *68*, 130. https://doi.org/10.15227/orgsyn.068.0130.

(11) Kauffman, G. B.; Fang, L. Y.; Viswanathan, N.; Townsend, G. Purification of Copper (i) Iodide; 2007; pp 101–103. https://doi.org/10.1002/9780470132531.ch20.

(12) Midya, S. P.; Subaramanian, M.; Babu, R.; Yadav, V.; Balaraman, E. Tandem Acceptorless Dehydrogenative Coupling–Decyanation under Nickel Catalysis. *The Journal of Organic Chemistry* **2021**, *86* (11), 7552–7562. https://doi.org/10.1021/acs.joc.1c00592.

(13) Iwasaki, T.; Miyata, Y.; Akimoto, R.; Fujii, Y.; Kuniyasu, H.; Kambe, N. Diarylrhodates as Promising Active Catalysts for the Arylation of Vinyl Ethers with Grignard Reagents.

*Journal of the American Chemical Society* **2014**, *136* (26), 9260–9263. https://doi.org/10.1021/ja5043534.
(14) Prasanna, R.; Guha, S.; Sekar, G. Proton-Coupled Electron Transfer: Transition-Metal-Free Selective Reduction of Chalcones and Alkynes Using Xanthate/Formic Acid. *Organic Letters* **2019**, *21* (8), 2650–2653. https://doi.org/10.1021/acs.orglett.9b00635.
(15) Zubar, V.; Sklyaruk, J.; Brzozowska, A.; Rueping, M. Chemoselective Hydrogenation of Alkynes to ( Z ) - Alkenes Using an Air-Stable Base Metal Catalyst. *Organic Letters* **2020**, *22* (14), 5423–5428. https://doi.org/10.1021/acs.orglett.0c01783.
(16) Lindlar, H. Ein Neuer Katalysator Für Selektive Hydrierungen. *Helvetica Chimica Acta* **1952**, *35* (2), 446–450. https://doi.org/10.1002/hlca.19520350205.
(17) Bamberger, N. D.; Ivie, J. A.; Parida, K.; McGrath, D. V.; Monti, O. L. A. Unsupervised Segmentation-Based Machine Learning as an Advanced Analysis Tool for Single Molecule Break Junction Data. *J. Phys. Chem. C* **2020**, *124* (33), 18302–18315. https://doi.org/10.1021/acs.jpcc.0c03612.
(18) Bamberger, N. D.; Dyer, D.; Parida, K. N.; McGrath, D. V.; Monti, O. L. A. Grid-Based Correlation Analysis to Identify Rare Quantum Transport Behaviors. *J. Phys. Chem. C* **2021**, *125* (33), 18297–18307. https://doi.org/10.1021/acs.jpcc.1c04794.
(19) Martín, S.; Grace, I.; Bryce, M. R.; Wang, C.; Jitchati, R.; Batsanov, A. S.; Higgins, S. J.; Lambert, C. J.; Nichols, R. J. Identifying Diversity in Nanoscale Electrical Break Junctions. *J. Am. Chem. Soc.* **2010**, *132* (26), 9157–9164. https://doi.org/10.1021/ja103327f.
(20) Hong, W.; Manrique, D. Z.; Moreno-García, P.; Gulcur, M.; Mishchenko, A.; Lambert, C. J.; Bryce, M. R.; Wandlowski, T. Single Molecular Conductance of Tolanes: Experimental and Theoretical Study on the Junction Evolution Dependent on the Anchoring Group. *J. Am. Chem. Soc.* **2012**, *134* (4), 2292–2304. https://doi.org/10.1021/ja209844r.
(21) Yanson, A. I.; Bollinger, G. R.; van den Brom, H. E.; Agraït, N.; van Ruitenbeek, J. M. Formation and Manipulation of a Metallic Wire of Single Gold Atoms. *Nature* **1998**, *395* (6704), 783–785. https://doi.org/10.1038/27405.
(22) Neese, F. The ORCA Program System. *WIREs Computational Molecular Science* **2012**, *2* (1), 73–78. https://doi.org/10.1002/wcms.81.
(23) Becke, A. D. Density-functional Thermochemistry. III. The Role of Exact Exchange. *J. Chem. Phys.* **1993**, *98* (7), 5648–5652. https://doi.org/10.1063/1.464913.
(24) Lee, C.; Yang, W.; Parr, R. G. Development of the Colle-Salvetti Correlation-Energy Formula into a Functional of the Electron Density. *Phys. Rev. B* **1988**, *37* (2), 785–789. https://doi.org/10.1103/PhysRevB.37.785.
(25) Ivie, J. A.; Bamberger, N. D.; Parida, K. N.; Shepard, S.; Dyer, D.; Saraiva-Souza, A.; Himmelhuber, R.; McGrath, D. V.; Smeu, M.; Monti, O. L. A. Correlated Energy-Level Alignment Effects Determine Substituent-Tuned Single-Molecule Conductance. *ACS Appl. Mater. Interfaces* **2021**, *13* (3), 4267–4277. https://doi.org/10.1021/acsami.0c19404.
(26) Venkataraman, L.; Park, Y. S.; Whalley, A. C.; Nuckolls, C.; Hybertsen, M. S.; Steigerwald, M. L. Electronics and Chemistry:  Varying Single-Molecule Junction Conductance Using Chemical Substituents. *Nano Lett.* **2007**, *7* (2), 502–506. https://doi.org/10.1021/nl062923j.
(27) Lo, W.-Y.; Bi, W.; Li, L.; Jung, I. H.; Yu, L. Edge-on Gating Effect in Molecular Wires. *Nano Lett.* **2015**, *15* (2), 958–962. https://doi.org/10.1021/nl503745b.
(28) Frisch, M. J.; Trucks, G. W.; Schlegel, H. B.; Scuseria, G. E.; Robb, M. A.; Cheeseman, J. R.; Scalmani, G.; Barone, V.; Petersson, G. A.; Nakatsuji, H.; Li, X.; Caricato, M.;

Marenich, A. V.; Bloino, J.; Janesko, B. G.; Gomperts, R.; Mennucci, B.; Hratchian, H. P.; Ortiz, J. V.; Izmaylov, A. F.; Sonnenberg, J. L.; Williams; Ding, F.; Lipparini, F.; Egidi, F.; Goings, J.; Peng, B.; Petrone, A.; Henderson, T.; Ranasinghe, D.; Zakrzewski, V. G.; Gao, J.; Rega, N.; Zheng, G.; Liang, W.; Hada, M.; Ehara, M.; Toyota, K.; Fukuda, R.; Hasegawa, J.; Ishida, M.; Nakajima, T.; Honda, Y.; Kitao, O.; Nakai, H.; Vreven, T.; Throssell, K.; Montgomery Jr., J. A.; Peralta, J. E.; Ogliaro, F.; Bearpark, M. J.; Heyd, J. J.; Brothers, E. N.; Kudin, K. N.; Staroverov, V. N.; Keith, T. A.; Kobayashi, R.; Normand, J.; Raghavachari, K.; Rendell, A. P.; Burant, J. C.; Iyengar, S. S.; Tomasi, J.; Cossi, M.; Millam, J. M.; Klene, M.; Adamo, C.; Cammi, R.; Ochterski, J. W.; Martin, R. L.; Morokuma, K.; Farkas, O.; Foresman, J. B.; Fox, D. J. *Gaussian 16 Rev. C.01*; Wallingford, CT, 2016.